\documentclass{aastex631}

\usepackage{graphicx}
\usepackage{dcolumn}
\usepackage{bm}
\usepackage{epsfig}
\usepackage{amsmath}
\usepackage{ulem}

\usepackage[T1]{fontenc} 
\usepackage[utf8]{inputenc} 
\usepackage[spanish, english]{babel} 

\begin{document}

\title{The Hubble Tension resolved by the DESI Baryon Acoustic Oscillations Measurements}

\author[0009-0009-3583-552X]{X. D. Jia}
\affiliation{School of Astronomy and Space Science, Nanjing University, Nanjing 210093, China}

\author[0000-0002-5819-5002]{J. P. Hu}
\affiliation{Ministry of Education Key Laboratory for Nonequilibrium Synthesis and Modulation of Condensed Matter, School of Physics, Xi’an Jiaotong University, Xi’an 710049, China}

\author[0000-0001-7176-8170]{D. H. Gao}
\affiliation{School of Astronomy and Space Science, Nanjing University, Nanjing 210093, China}

\author[0000-0003-0672-5646]{S. X. Yi}
\affiliation{School of Physics and Physical Engineering, Qufu Normal University, Qufu 273165, China}

\author[0000-0003-4157-7714]{F. Y. Wang}
\affiliation{School of Astronomy and Space Science, Nanjing University, Nanjing 210093, China}
\affiliation{Key Laboratory of Modern Astronomy and Astrophysics (Nanjing University), Ministry of Education, China}

\correspondingauthor{F. Y. Wang}
\email{fayinwang@nju.edu.cn}

\begin{abstract}
The $\Lambda$ cold dark matter ($\Lambda$CDM) cosmological model provides a good description of a wide range of astrophysical and cosmological observations. However, severe challenges to the phenomenological $\Lambda$CDM model have emerged recently, including the Hubble constant tension and the significant deviation from the $\Lambda$CDM model reported by the Dark Energy Spectroscopic Instrument (DESI) collaboration. Despite many explanations for the two challenges have been proposed, the origins of them are still intriguing mysteries. Here, we investigate the DESI Baryon Acoustic Oscillations (BAOs) measurements to interpret the Hubble constant tension. Employing a non-parametric method, we find that the dark energy equation of state $w(z)$ evolves with redshift from DESI BAO data and type Ia supernovae. From the Friedmann equations, the Hubble constant ($H_0$) is derived from $w(z)$ model-independently. We find that the values of $H_0$ show a descending trend as a function of redshift, and can effectively resolve the Hubble constant tension. Our study finds that the two unexpected challenges to the $\Lambda$CDM model can be understood in one physical framework, e.g., dynamical dark energy.

\end{abstract}

\keywords{Hubble constant(758) --- Dark energy(351) --- Type Ia supernovae(1728) --- Baryon acoustic oscillations(138) }

\section{Introduction}\label{sec:intro}
The $\Lambda$ cold dark matter ($\Lambda$CDM) model is considered as the standard cosmological model \citep{Peebles2003}. It accounts for the observed cosmic expansion, the cosmic microwave background
(CMB), the dynamics of galaxies, the statistical properties of large scale structures of the
Universe  and the observed abundances of light nuclei. Despite its unprecedented successes, fundamental challenges for the $\Lambda$CDM model have emerged in recent years \citep{Perivolaropoulos2022,2022JHEAp..34...49A}. The most crucial challenges are the Hubble constant tension \citep{2019NatAs...3..891V,DiValentino2021} and the deviation from the $\Lambda$CDM model reported by the Dark Energy Spectroscopic Instrument (DESI) collaboration \citep{Adame2025,DESI_DR2}. 

Fitting the standard $\Lambda$CDM model to the CMB anisotropies measurements from $z \sim 1100$, the Hubble constant $H_0 = 67.4 \pm 0.5$ km s$^{-1}$ Mpc$^{-1}$ with 1$\sigma$ confidence level is derived by the Planck Collaboration \citep{Planck_2020}.  In contrast, the SH0ES team measured the Hubble constant $H_0 = 73.04\pm 1.04$ km s$^{-1}$ Mpc$^{-1}$ through observations of Cepheid variables and type Ia supernovae (SNe Ia) \citep{Riess2022}. The discrepancy between early-universe and late-universe measurements of $H_0$, exceeding $5 \sigma$ significance, is termed as the Hubble constant tension \citep{2019NatAs...3..891V,Riess2020}. 
Various attempts have been proposed to explain this tension, which can be broadly classified into three categories \citep{DiValentino2021,2025PDU....4901965D}: early-universe modifications, late-universe effects, and local-universe phenomena. For the early-universe approach, it attempts to modify the pre-recombination physics in order to alter the sound horizon scale, thereby increasing the value of $H_0$ inferred from CMB data \citep{2019PhRvL.122v1301P,2020PhRvL.125r1302J}. For the late-universe solutions, they aim to modify the cosmic expansion history through specific dark energy parameterizations, thereby increasing the $H_0$ value at low redshift to be consistent with local measurements \citep{2017NatAs...1..627Z,Benevento2020,2025arXiv250310806C}. For the local universe phenomena, these models attempt to explain the larger local $H_0$ through local either inhomogeneities \citep{2025MNRAS.536.3232M,2025MNRAS.540..545B} or unaccounted systematic errors in the local distance ladder \citep{Freedman2021,Mortsell2022}. 

On the other hand, the baryon acoustic oscillation (BAO) measurements from the DESI collaboration demonstrate a clear preference for dynamical dark energy model over the standard $\Lambda$CDM model \citep{Adame2025,DESI_DR2}. By combining different SNe Ia samples, the preference for the dynamical dark energy model over the standard $\Lambda$CDM model ranges from $2.8$-$4.2\sigma$ \citep{DESI_DR2}. 
Under the Chevallier-Polarski-Linder (CPL) parametrization \citep{Chevallier2001,Linder2003}, it exhibits distinct dynamical behavior, demonstrating a deviation from the $\Lambda$CDM model exceeding $3.1 \sigma$ even without including SNe Ia sample \citep{DESI_DR2,DESI_Dark_energy}. Using a non-parametric method, the dark energy equation of state $w(z)$ is found to be evolving with redshift \citep{DESI_DR2,DESI_Dark_energy,Gu2025}, which is robust and stable for different
model choices. 
The DESI BAO data has sparked widespread discussion \citep{2024ApJ...976....1L,2024PhRvD.110l3533P,2025MNRAS.542.1063H,2025arXiv250107361L,2025arXiv250522369L,2025MNRAS.542L..24C,2025PhRvD.111l3504P,2025arXiv250103480P,2025PhRvL.134r1002Y,2026JHEAp..4900428O}.
Here, we interpret the Hubble tension using the DESI DR2 BAO measurements.

The CPL parametrization effectively describes dark energy under a ad-hoc functional form. In order to study $w(z)$ model-independently, we use a non-parametric approach \citep{Huterer2005,Riess2007,DESI_Dark_energy}. This method is widely used, which enables the comparison of different redshift bins without assuming any functional form \citep{Jia2022,DESI_Dark_energy}. The fundamental assumption of the binning method is that the physical quantity can be approximated as constant within a redshift interval. 
If no evolution exists, the results will remain consistent across all redshift intervals. If an evolutionary trend is present, it helps quantify deviations from the standard cosmological model.

In this letter, we investigate the evolution of the dark energy equation of state using a non-parametric approach and derive the expression for $H_0$ starting from the Friedmann equations. The Letter is organized as follows. In Section \ref{sec:method}, we introduce the data and methods. The evidence of evolving dark energy is shown in Section \ref{sec:The evidence of evolving dark energy}. In Section \ref{sec:Calculating $H_0$ from evolving dark energy}, we calculate the $H_0$ from the evolving dark energy. The summary is given in Section \ref{sec:summary}.

\section{Method}\label{sec:method}
\subsection{Datasets}
Recently, the second round of cosmological constraints based on BAO from DESI collaboration has been released \citep{DESI_DR2}. The results of DESI DR2 BAO demonstrate a preference for dynamical dark energy over the standard cosmological model. When combined with other cosmological observations (e.g., CMB, SNe Ia), this preference exceeds $2.8 \sigma$ significance \citep{DESI_DR2}. The BAO measurements presented in Table IV from DESI DR2 measurements \citep{DESI_DR2} contain the transverse comoving distance and Hubble expansion rate at specified redshifts. The measurement in the direction transverse to the line-of-sight gives the comoving angular diameter distance
\begin{equation}
    D_M(z) = \frac{c}{H_0} \int_0^z \frac{H_0}{H(z^\prime)} dz^\prime.
\end{equation}
The measurement in the direction along the line-of-sight is sensitive to the expansion rate as 
\begin{equation}
    D_H(z) = \frac{c}{H(z)}.
\end{equation}
However, it is evident that these two distances are correlated, and their covariance matrix must be considered in the analysis. The sound horizon is typically determined from CMB data with
\begin{equation}
    r_\mathbf{d} = \int_{z_d}^\infty \frac{c_s(z)}{H(z)} dz. 
\end{equation}
Here $c_s(z)$ is the speed of sound in the photon-baryon fluid, and $z_d \approx 1089$ denotes the redshift at which photons and baryons decouple \citep{Planck_2020}. Under the standard pre-recombination physics, it can be calculated as 
\begin{equation}
\begin{split}
    r_{\mathrm{d}}=  147.05~ \mathrm{Mpc} &\times   \left(\frac{\omega_{\mathrm{b}}}{0.02236}\right)^{-0.13} \times \left(\frac{\omega_{\mathrm{bc}}}{0.1432}\right)^{-0.23} \\
    & \times \left(\frac{N_{\mathrm{eff}}}{3.04}\right)^{-0.1}.
\end{split}
\end{equation}
Here, $\omega_\mathrm{b} = \Omega_b h^2$ and $\omega_{\mathrm{bc}} = (\Omega_\mathrm{b} + \Omega_\mathrm{c}) h^2$. They represent the density of baryons and cold dark matter, respectively. The parameter $N_{\mathrm{eff}}$ is the effective number of relativistic degrees of freedom. The value of these parameters are adopted from the fitting results of $Planck$ \citep{Planck_2020}. 
\par

\par
The uniform intrinsic luminosity makes SNe Ia a standard candle, which is crucial for measuring the Hubble constant especially at low redshifts. At present, the three most widely used SNe Ia samples are, Pantheon plus \citep{Pantheon+}, DESY5 \citep{DESY5}, and Union3 \citep{Union3}. They are commonly combined with other cosmological probes for joint constraints \citep{DESI_DR2}. The significance for rejecting the $\Lambda$CDM model is $2.8 \sigma$, $3.8 \sigma$, and $4.2 \sigma$ when combined with DESI, CMB, and the Pantheon plus/Union3/DESY5 SNe samples, respectively. Considering that the Union3 dataset has not yet been fully released, conducting the binning method with it remains hard. Moreover, since its deviation from the $\Lambda$CDM model is intermediate between the other two results, we exclude this dataset from our analysis. By standardizing SNe Ia light curves, their fitting parameters can be used to estimate the apparent magnitude. Using the local distance ladder, the SH0ES team calibrated the absolute magnitudes of SNe Ia \citep{Riess2022}. Based on their results, we can calculate the absolute magnitudes from their apparent magnitudes. When fitting SNe Ia data, it is essential to account for both statistical uncertainties, and the covariance matrix of systematic errors. An important consideration is that low-redshift data are particularly susceptible to peculiar velocity effects. To avoid this bias, we exclude all SNe Ia with $z < 0.01$.

\subsection{Non-parametric method}
The CPL parametrization effectively describes dark energy under a predetermined functional form, but this approach inherently assumes the model's validity. In order to focus on the quantity of $w(z)$ rather than estimating the parameters of an assumed functional form, we use a non-parametric approach. The binning method is widely used in cosmology, which enables the comparison of different redshift intervals without assuming any functional form \citep{Huterer2005,2017NatAs...1..627Z,2025arXiv250207185B}. 
\par
The fundamental assumption of the binning method is that the physical quantity can be approximated as constant within a redshift interval. Here, our focus is on the equation of state of dark energy $w(z)$. The binning method can explore potential evolutionary trends in dark energy by introducing additional degrees of freedom. If no evolution exists, the results will remain consistent across all redshift intervals. If an evolutionary trend is present, it helps quantify deviations from the standard cosmological model. From the Friedmann equation, the expansion rate in a flat universe can be expressed as
\begin{equation}
    H^2(z) = H_0^2 \left[\Omega_{\mathrm{m}} (1+z)^3 + \Omega_{\mathrm{DE}} f_{\mathrm{DE}}(z) \right], \label{eq_H(z)_wcdm}
\end{equation}
where $f_{\mathrm{DE}}(z)=\exp \left(3 \int\left(\mathrm{~d} z^{\prime} / 1+z^{\prime}\right)\left[1+w\left(z^{\prime}\right)\right]\right)$. The equation of state $w(z)$ is defined as $w(z) = P_{\mathrm{DE}}(z)/\rho_{\mathrm{DE}}(z)$. Assuming that the dark energy equation of state $w(z)$ can be approximated as a constant within a given redshift interval, it can be expressed as:
\begin{equation}
    w(z) = w_i, \qquad z_{i-1} < z \leq z_i.
\end{equation}
The parameter $z_i$ is the upper boundary of redshift bin, and $z_0 = 0$. For $w(z) = -1$, it describes the $\Lambda$CDM model. Under the assumption that $w(z)$ remains constant within each redshift bin, the function $f_{\mathrm{DE}}(z)$ becomes a piece-wise function of the form
\begin{equation}
    f_{\mathrm{DE}}\left(z_{n-1}<z \leq z_{n}\right)=(1+z)^{3\left(1+w_{n}\right)} \prod_{i=0}^{n-1}\left(1+z_{i}\right)^{3\left(w_{i}-w_{i+1}\right)},
\end{equation}
where $w_i$ is the value of $w(z)$ in the $i$th redshift bin, and $n$ is the total number of the redshift bin. The value of $w_0$ can be assumed to be a constant. Given that $z_0 = 0$ under this condition, it follows that the value of $w_0$ does not affect the result. Under the piece-wise function approach, we derive its redshift evolution across distinct bins. 

\par
Since both the luminosity distance and Hubble expansion rate involve integrals and summations over different redshift bins, which introduce inherent correlations among the fitted $w_i$ values \citep{Riess2007}. In order to remove these correlations, we used the component analysis by diagonalizing the covariance matrix \citep{Huterer2005,Riess2007}. The covariance matrix of the constrained $w_i$ can be generated by taking the average over the chain, i.e., 
\begin{equation}
\mathbf{C}=\langle \mathbf{H} \mathbf{H}^{\rm T}\rangle-\langle\mathbf{H}\rangle\langle\mathbf{H}^{\rm T}\rangle,
\end{equation}
where $\mathbf{H}$ is a vector with components $w_i$, and the angle brackets denote the average value of the vector $\mathbf{H}$. Through the Fisher matrix,we can find a set of uncorrelated basis that diagonalize this covariance matrix. It can be expressed as
\begin{equation}
\mathbf{F} \equiv \mathbf{C}^{-1} \equiv \mathbf{O}^{\mathrm{T}} {\Lambda} \mathbf{O} ,
\end{equation}
\begin{equation}
\mathbf{T}=\mathbf{O}^{\mathrm{T}}  {\Lambda}^{\frac{1}{2}} \mathbf{O} .
\end{equation}
Here, $\Lambda$ is the diagonalized covariance for transformed bins, $\mathbf{T}$ is the transformation matrix. Finally, the results are uncorrelated after multiplying the transformation matrix as 
\begin{equation}
\tilde{\mathbf{H}}=\mathbf{TH}.
\end{equation}
The result exhibits a diagonal covariance matrix $\Lambda^{-1}$, demonstrating it is uncorrelated.
\par
For parameter estimations, we employ the Markov Chain Monte Carlo (MCMC) code $emcee$ \citep{2013PASP..125..306F}. The prior of $w_i$ is chosen as $w_i \in$ [-5, 1], wider than that adopted by the DESI collaboration \citep{DESI_Dark_energy}. The marginalized posterior distributions of $\Omega_m$ show slight discrepancy between DESI BAO and different SNe Ia samples \citep{DESI_DR2}. While these differences remain within the $3 \sigma$ confidence level, we adopt a broad prior $\Omega_m \in$ [0.25, 0.45], which fully encompasses the posterior ranges from literature. The prior of $H_0\in$ [60, 80] km s$^{-1}$ Mpc$^{-1}$ is adopted. The full parameters \{$\Omega_M, H_0, w_1, ..., w_i$\} are fitted simultaneously. 

The posterior distributions of the free parameters are determined through chi-square fitting, which can be expressed as:
\begin{equation}
    \chi_{\theta}^{2}=\chi_{\mathrm{BAO}}^2+\chi_{\mathrm{SNe \ Ia}}^2.
\end{equation}
The value of $\chi^2_{BAO}$ is 
\begin{equation}
\chi_{BAO}^{2}= \left[\nu_{\mathrm{obs}}\left(z_{i}\right)-\nu_{\mathrm{th}}\left(z_{i}\right)\right]\mathbf{C_{BAO}}^{-1}\left[\nu_{\mathrm{obs}}\left(z_{i}\right)-\nu_{\mathrm{th}}\left(z_{i}\right)\right]^{T},
\end{equation}
where $\nu_{obs}$ is the vector of the BAO measurements at each redshift $z$ (i.e. $D_V\left(r_{s,{\rm fid}}/r_s\right), D_M/r_s, D_H/r_s$).
The value of $\chi_{\mathrm{SNe \ Ia}}^2$ is 
\begin{equation}
\chi_{\mathrm{SNe \ Ia}}^{2}= \left[\mu_{\mathrm{obs}}\left(z_{i}\right)-\mu_{\mathrm{th}}\left(z_{i}\right)\right]\mathbf{C_{SNe \ Ia}}^{-1}\left[\mu_{\mathrm{obs}}\left(z_{i}\right)-\mu_{\mathrm{th}}\left(z_{i}\right)\right]^{T}
.\end{equation}
The parameter $\mu_{\mathrm{obs}}(z_i)$ is the distance module, which is calibrated using the absolute magnitude derived from the local distance ladder \citep{Riess2022}. 

\section{The evidence of evolving dark energy}\label{sec:The evidence of evolving dark energy}
\begin{figure}
    \centering
    \includegraphics[width=0.6\textwidth]{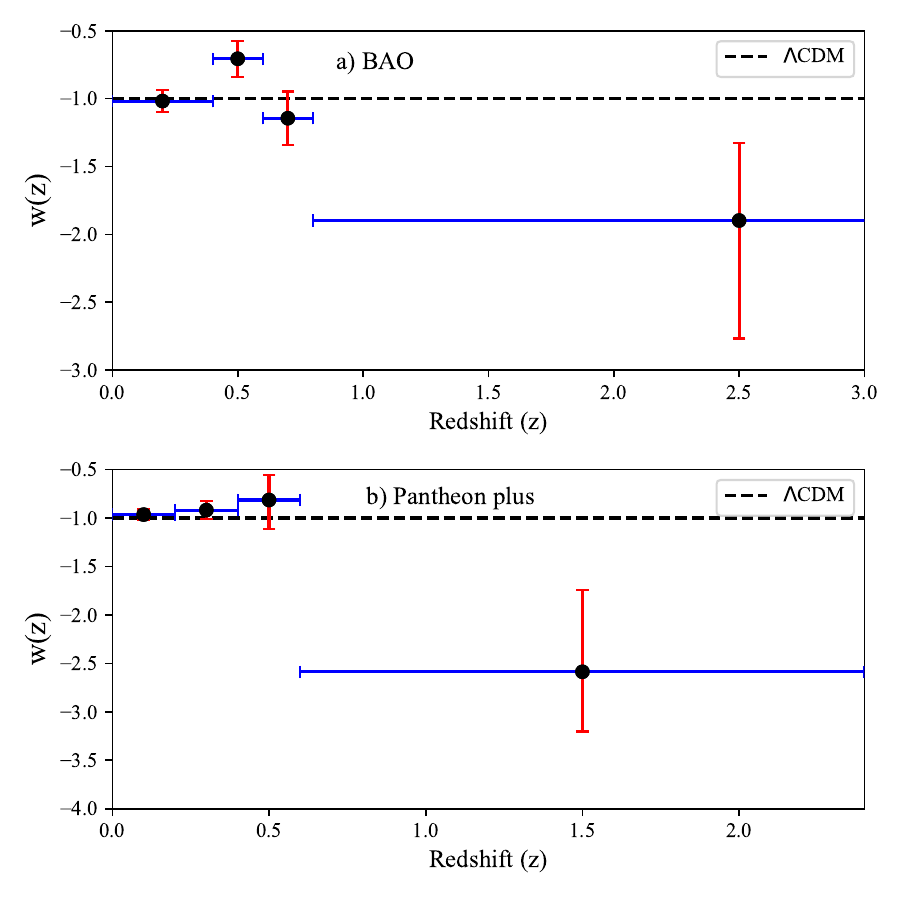} 
    \caption{Fitting results of the equation of state $w(z)$ of dark energy with the non-parametric method. Panel (a). The black points represent the best-fit values of $w(z)$ with $1\sigma$ confidence level (red bar) at different redshift bins using the DESI DR2 BAO measurements. The black dashed line represents the standard $\Lambda$CDM model with $w(z)=-1$. The redshift intervals of the bins are shown as the blue bars. A descending trend of $w(z)$ is found. Panel (b). Same as panel (a), but for Pantheon plus SNe Ia sample. The $w(z)$ values from BAO and SNe Ia are consistent within 1$\sigma$ confidence level. 
    }
    \label{F_w_D+P}
\end{figure}

We systematically analyze the potential impacts of various factors on our fitting procedure. To avoid potential biases introduced by joint analysis, we conduct independent analyses on each sub-sample. For individual subsamples, due to their limited constraining power, we have to adjust the redshift binning scheme to achieve stronger constraints. For the DESI DR2 BAO measurements, we group the data according to its coverage, with the following upper boundaries at $z_i = 0.4, 0.6, 0.8$ and $4.2$. 
We adopt four redshift bins in the range $0<z<4.2$, which is similar to the binning scheme for different tracers as in the DESI DR2 BAO measurements \citep{DESI_DR2}. The best-fit result of $w_i$ is shown in the fourth column of the Table \ref{T} and the panel (a) in the Figure \ref{F_w_D+P}, which shows a descending trend with redshift. The result exhibits a consistent descending trend, regardless of whether the sound horizon scale is calibrated with or without CMB data. We also test the descending trend using two SNe Ia samples, such as Panthon plus \citep{Scolnic2022} and DESY5 \citep{DESCollaboration2024}. Both the two samples give a similar descending trend. 
The constraints on $w(z)$ from Pantheon plus sample are shown in panel (b) of Figure \ref{F_w_D+P}. The upper boundaries are set as $z_i = 0.2, 0.4, 0.6$ and $2.4$. The DESY5 results are similar to those from Pantheon plus, so we omit them here for brevity.
We attempt to let $\Omega_m$ vary freely, but this approach failed to provide effective constraints on $w(z)$ at high redshifts. Consequently, we adopt a fixed prior range for $\Omega_m$, derived from CMB data \citep{Planck_2020}. We test different values of $\Omega_m$. The results show that while $\Omega_m$ affects the absolute values of the fitted $w(z)$, it does not alter the evolutionary trend. 
The values from different datasets (BAO and SNe Ia) are consistent within 1$\sigma$ confidence level. Therefore, the descending trend of $w(z)$ is robust from DESI DR2 BAO measurements and SNe Ia. 

\begin{figure}
    \centering
    \includegraphics[width=0.6\textwidth]{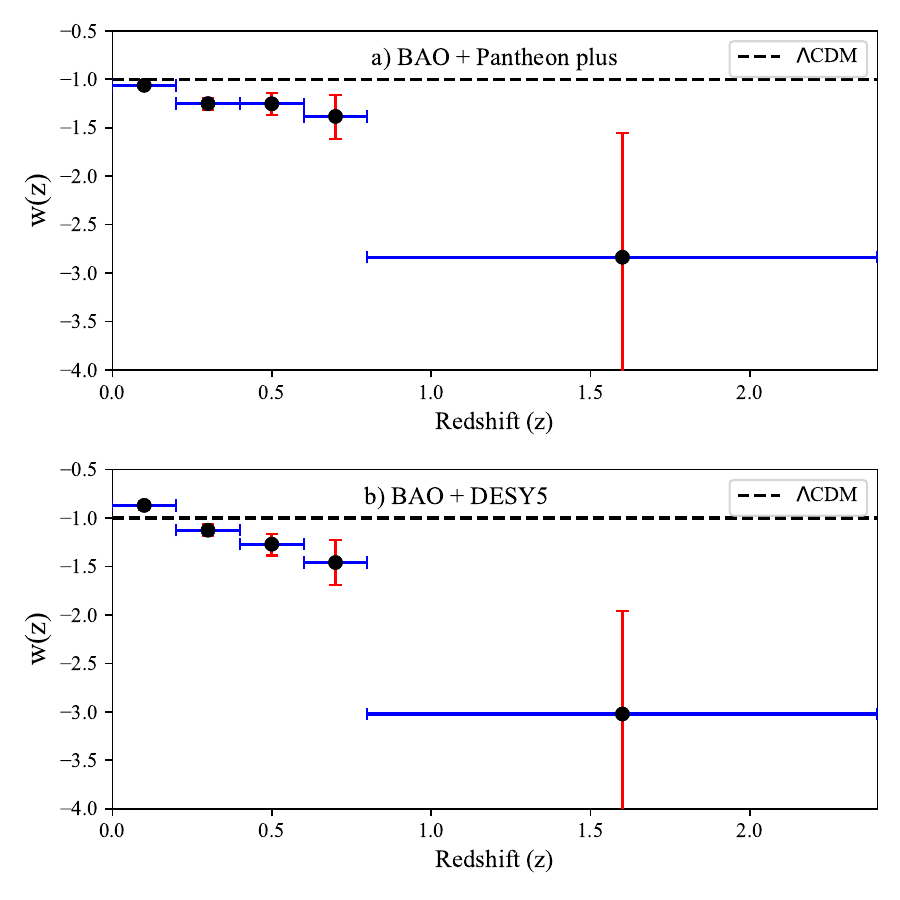} 
    \caption{Fitting results of the equation of state $w(z)$ of dark energy with the non-parametric method using BAO and SNe Ia. Panel (a). The black points represent the best-fit values of $w(z)$ with $1\sigma$ confidence level (red bar) at different redshift bins using the DESI DR2 BAO measurements and Pantheon plus SNe Ia sample. The black dashed line represents the standard $\Lambda$CDM model with $w(z)=-1$. The redshift intervals of the bins are shown in the blue bars. A descending trend of $w(z)$ is found. It crosses $w=-1$, suggesting quintom-like behavior. Panel (b). Same as panel (a), but for the DESI DR2 BAO measurements and DESY5 SNe Ia sample. The $w(z)$ values in panels (a) and (b) are consistent within 1$\sigma$ confidence level.}
    \label{F_w_DP+DD}
\end{figure}

\begin{figure*}
    \centering
    \includegraphics[width=0.5\textwidth,angle=0]{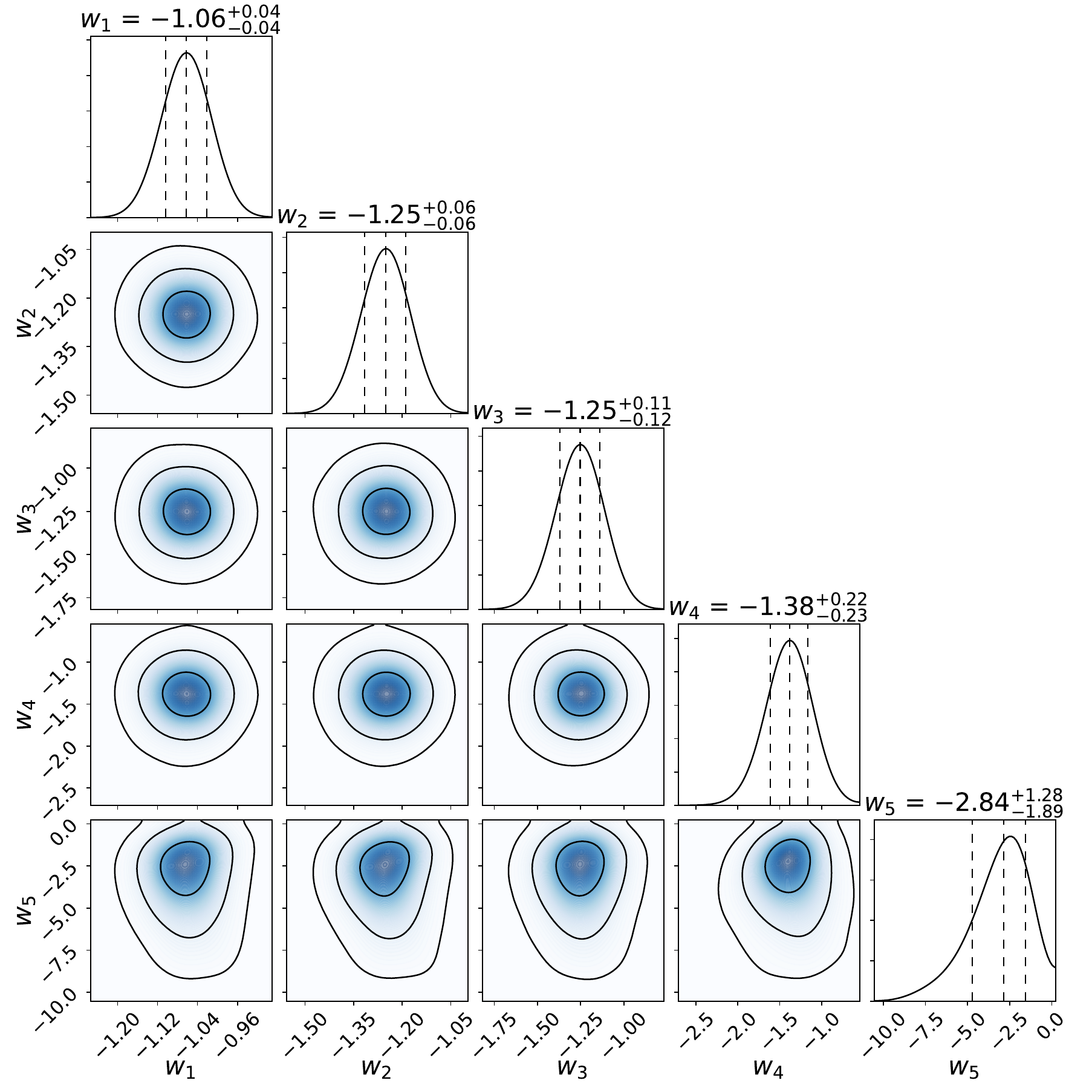}
    \caption{Corner plot of $w_i$ from the DESI DR2 BAO + Pantheon plus sample. The panels on the diagonal show the 1D histogram for each parameter obtained by marginalizing over the other parameters. The off-diagonal panels show two-dimensional projections of the posterior probability distributions for each pair of parameters, with contours to indicate $1\sigma-3\sigma$ confidence levels.}
    \label{F_cor_DESI+Pan}
\end{figure*}

\begin{figure*}
    \centering
    \includegraphics[width=0.5\textwidth,angle=0]{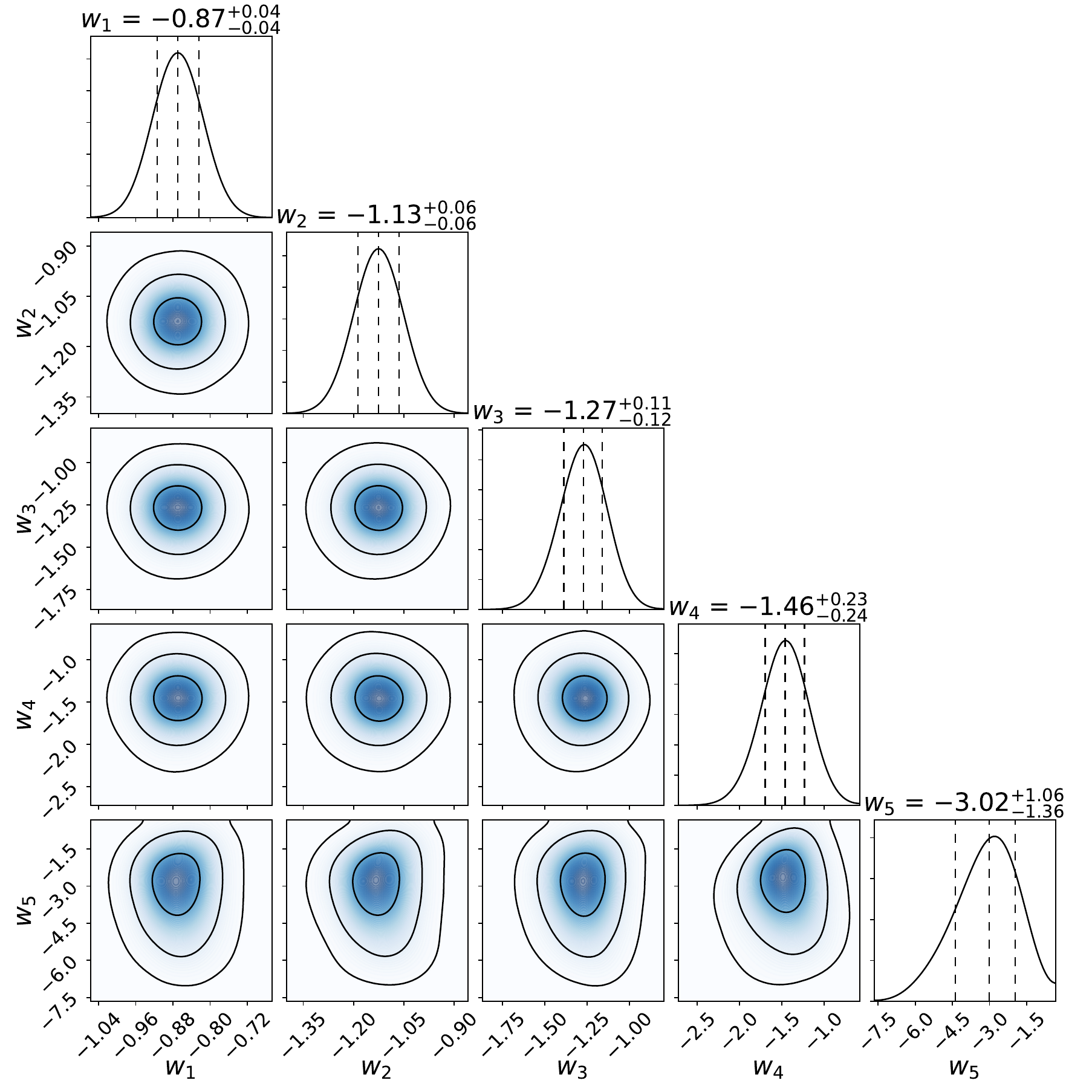}
    \caption{Same as Figure \ref{F_cor_DESI+Pan}, but for the DESI DR2 BAO + DESY5 SNe Ia sample.}
    \label{F_cor_DESI+DESY5}
\end{figure*}

In order to obtain tight constraints on the equation of state $w(z)$ of dark energy, we combine the DESI DR2 BAO measurements and SNe Ia data. We systematically investigate different sample combinations and binning methods to assess their impacts on the results. The consistently descending trends are shown across different binning methods, indicating the robustness of the evolutionary behavior in the equation of state parameter. Here, we present the results of dividing the sample into five redshift bins with upper boundaries at $z_i = 0.2, 0.4, 0.6, 0.8$ and $4.2$. The SNe Ia sample contains limited high-redshift data, making the BAO measurements decisive for determining the upper bound at $z = 4.2$. The best-fit values of $w_i$ with 1$\sigma$ error bars from DESI DR2 BAO+Pantheon plus sample and DESI DR2 BAO+DESY5 sample, are shown in panels (a) and (b) in Figure \ref{F_w_DP+DD}, respectively. The corner plot of these results are shown in Figures \ref{F_cor_DESI+Pan} and \ref{F_cor_DESI+DESY5}. Both distinct sample sets reveal a consistent trend for the evolution behavior of $w(z)$. The values are consistent in $1 \sigma$ confidence level, as shown in Table \ref{T}. Despite slight variation in the absolute values of $w(z)$, all datasets consistently demonstrate the same descending trend. 

Notably, the current SNe Ia samples lack sufficient high-redshift data, resulting in weak constraints at $z > 1$. To mitigate this, we have to adopt a broader redshift binning strategy. During the fitting procedure, we simultaneously allow both $\Omega_m$ and $H_0$ to vary as free parameters. The two different sample combinations yield consistent $\Omega_m$ results, agreeing within $1 \sigma$ confidence level. However, the DESI DR2 BAO and Pantheon plus SNe Ia sample produce a larger fitted $H_0$ value. It likely led to its smaller $w(z)$ value in the first redshift bin. At low redshifts, the constraints are tight on $w(z)$, but degrade significantly in the high redshift bin due to the limited data number. DESI collaboration also found consistent redshift evolution of $w(z)$ trend for a broad range of parametric and non-parametric methods using the DESI DR2 BAO measurements, Planck CMB data, and three SNe Ia samples \citep{DESI_Dark_energy}. Our results are in good agreement with the finding of DESI collaboration. Both of the different sample combinations yield results in agreement, revealing a descending trend. It is worth noting that there is a slight difference between the two results. Specifically, for the DESI DR2 BAO and Pantheon plus SNe Ia sample at low redshift, the values of $w(z)$ do not cross $-1$. In our fitting procedure, we set paremeters $w(z)$, $\Omega_m$ and $H_0$ as free parameters. The DESI DR2 BAO and Pantheon plus SNe Ia sample obtain a high value of $H_0$, which is likely the reason why its constraints on $w(z)$ do not cross the boundary of $w = -1$.

Several different binning methods for $w(z)$ were tested, including logarithmic binning, uniform number binning and Bayesian block methodology \citep{Jia2025}. Consistent trends of $w(z)$ were found across different binning approaches. This excludes the possibility that the evolutionary trend is a statistical artifact introduced by the binning strategy. 
Overall, the $w_i$ values from the DESI DR2 BAO measurements combining different SNe Ia samples are consistent in 1$\sigma$ confidence level. This supports that the redshift evolution of $w(z)$ is robust and not caused by biases inherent to any particular sample. Our findings agree well with those from other approaches \citep{DESI_Dark_energy,Gu2025,Li2025}. Therefore, there exists a redshift-dependent evolutionary trend in the equation of state $w(z)$ of dark energy from the DESI DR2 BAO measurements and SNe Ia.

\section{Calculating $H_0$ from evolving dark energy}\label{sec:Calculating $H_0$ from evolving dark energy}
In this section, we derive the Hubble constant ($H_0$) using the $w_i$ measurements from the Friedmann equations.
Within the Friedmann-Lemaitre-Robertson-Walker metric, the Friedmann equations are 
\begin{equation}
    H^{2}=\left(\frac{\dot{a}}{a}\right)^{2}=\frac{8 \pi G}{3} \rho, \label{Friedmann_1}
\end{equation}
\begin{equation}
    \frac{\ddot{a}}{a}=-\frac{4 \pi G}{3}(\rho+3 p), \label{Friedmann_2}
\end{equation}
where $a=1/(1+z)$ is the scale factor, and $\dot{a}=da/dt$.
Combining equations (\ref{Friedmann_1}) and (\ref{Friedmann_2}), we find
\begin{equation}
    \rho=\rho_{0} \exp \left(-3 \int_{1}^{a}\left(1+w\left(a^{\prime}\right)\right) \frac{d a^{\prime}}{a^{\prime}}\right). \label{rho}
\end{equation}
The parameter $\rho_0$ denotes the cosmic total density at present. Substituting equation (\ref{rho}) into (\ref{Friedmann_1}), we find 
\begin{equation}
    H_{0}^{2}=H(z)^{2} \times \frac{\rho_c}{\rho_0} (1+z)^{-3} \exp \left(-3 \int_{0}^{z} w\left(z^{\prime}\right) \frac{d z^{\prime}}{1+z^{\prime}}\right), \label{eq_H0z}
\end{equation}
where $\rho_c=3H_0^2/8\pi G$ is the critical density of the universe. This equation gives the relation between the $w(z)$ evolution and the resulting behavior of $H_0$. 
The above derivation only depends on the cosmological principle and Einstein's equations.

\begin{figure}
    \centering
    \includegraphics[width=0.6\textwidth]{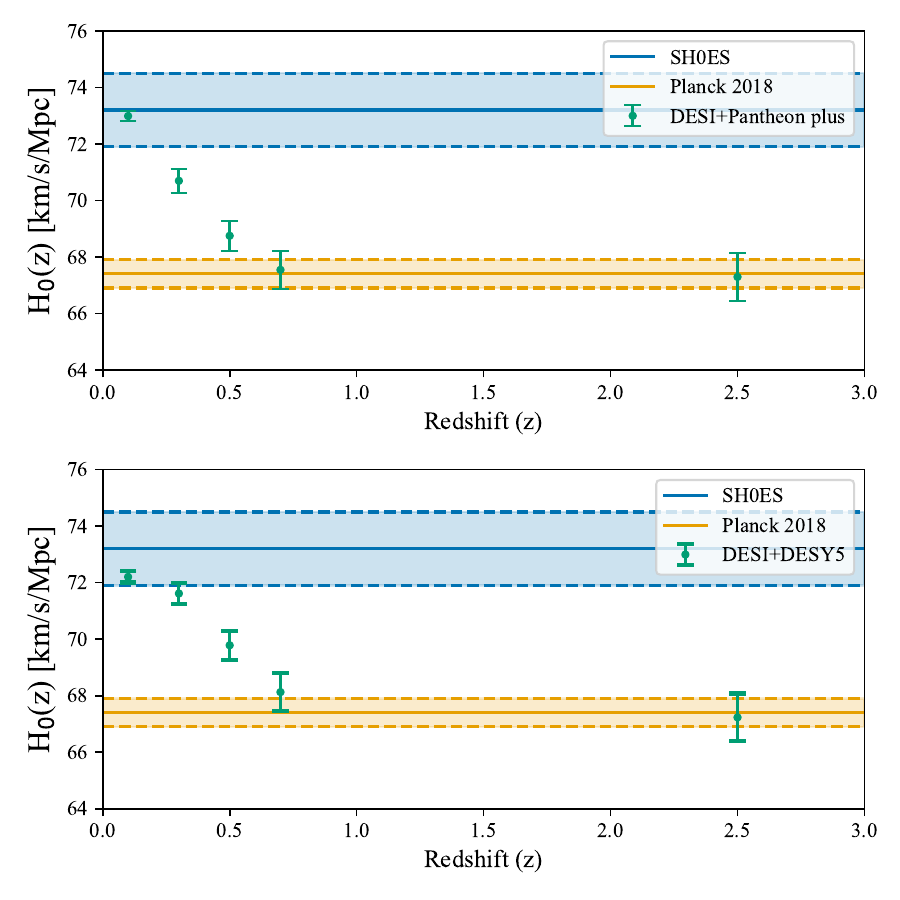} 
    \caption{The descending trend of the Hubble constant $H_0$ derived from the dark energy equation of state $w(z)$ in Figure \ref{F_w_DP+DD}. Panel (a). The green points indicate the maximum a posteriori estimates of $H_0(z)$ with $1 \sigma$ error bars from the DESI DR2 BAO measurements and Pantheon plus SNe Ia sample. The redshift of the green points correspond to the midpoint of each redshift bin. The blue solid line and blue band show the $H_0$ value and $1\sigma$ confidence level measured from the local distance ladder \citep{Riess2022}. The orange solid line and orange band represent the $H_0$ value and $1\sigma$ confidence level derived from the CMB data \citep{Planck_2020}. The $H_0$ value agrees with that measured from the local distance ladder in $1\sigma$ confidence level at low redshift, and it gradually drops to the value measured from the Planck CMB at high redshift. The Hubble tension is effectively resolved by DESI DR2 BAO measurements. Panel (b). Same as panel (a), but for the DESI DR2 BAO measurements and DESY5 SNe Ia sample. The $H_0$ values in panels (a) and (b) are consistent within 1$\sigma$ confidence level.}
    \label{F_H0_DPDD}
\end{figure}

From equation (\ref{eq_H0z}), we calculate the $H_0$ values from above $w_i$ measurements in different redshift bins, as $H_{0, i}$. It forms a piece-wise function of $H_0(z)$, denotes the value of $H_{0, i}$ in the $i$-th redshift bin. It should be noted that $H_0(z)$ represents the value determined from the observation at $z$ by extrapolating $H(z)$ from higher $z$ to $z = 0$, by assuming a background cosmological model. For the first term in the right hand of equation (\ref{eq_H0z}) known as the cosmic expansion rate, we can derive it from equation (\ref{eq_H(z)_wcdm}) using the best-fit value of $w_i$ from DESI DR2 BAO and Pantheon plus SNe Ia sample. 
The second term in the right hand of equation (\ref{eq_H0z}) represents the background cosmological model.
We choose the standard $\Lambda$CDM model given by Planck CMB data as the background model, considering that the $H_0$ value from CMB is measured in this model \citep{Planck_2020}. Such an evolving $w(z)$ would consequently induce corresponding variations in estimates of $H_0(z)$ derived under the standard cosmological model as
\begin{equation}
    H_{0}^{2}(z)= \frac{H(z)^{2}_{w(z)}}{\Omega_m (1+z)^3 + \Omega_{\Lambda}} .\label{eq_H0_w/LCDM}
\end{equation}
The parameter $H(z)_{w(z)}$ means the value of $H(z)$ calculated from combining the results of $w(z)$ with equation (\ref{eq_H0z}). The parameters in the denominator are taken from the background model. More details can be found in \cite{2021PhRvD.103j3509K}. The results from the DESI DR2 BAO combined with Pantheon plus SNe Ia sample are shown in panel (a) of Figure \ref{F_H0_DPDD}. In panel (b), $H_0(z)$ values are derived from the DESI DR2 BAO combined with DESY5 SNe Ia sample. The $x$-axis represents the midpoint of each redshift bin, while the $y$-axis shows the $H_0$ values derived from the best-fit results. From this figure, we can see that $H_0(z)$ shows a descending trend with redshift in both cases. The values of $H_0(z)$ are consistent in $1\sigma$ confidence level for the two panels. Notably, the $H_0(z)$ value agrees with that measured from the local distance ladder in $1\sigma$ confidence level at low redshift, and it gradually drops to the value measured from the CMB at high redshift, as shown in Table \ref{T_H}. The $\Lambda$CDM model and the $w(z)$  model have distinct cosmic expansion histories. If the true $w(z)$ is varying, attempting to describe it within the $\Lambda$CDM model framework will lead to its dynamics being erroneously interpreted as variations in other parameters, such as $H_0$. Here, we adopt a non-parametric, model-independent methodology. The $w(z)$ values are derived directly from the observational data, without relying on any model priors. As our analysis shows, mapping the evolution of $w(z)$ to the $H_0(z)$ parameter produces a result that aligns with the local distance ladder at low redshifts and CMB at high redshifts. Therefore, the Hubble tension can be effectively resolved by the DESI DR2 BAO measurements and SNe Ia sample. 

\begin{figure}
    \centering
    \includegraphics[width=0.5\textwidth]{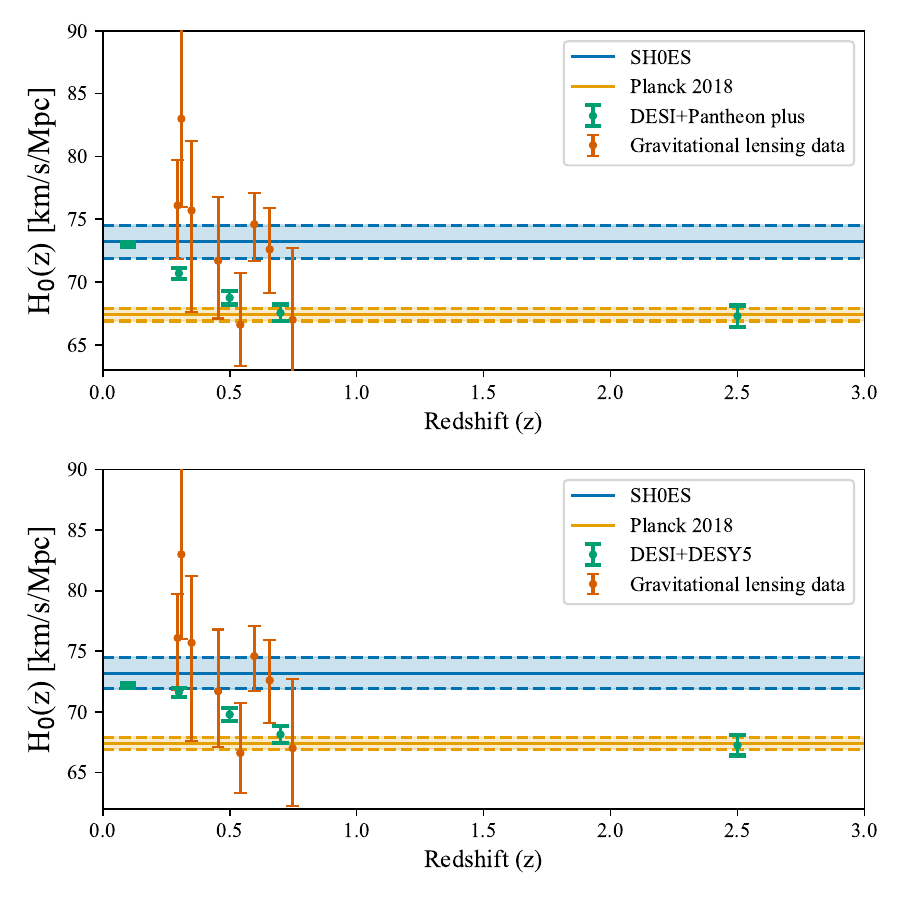}
    \caption{The descending trend of the Hubble constant $H_0$ derived from the dark energy equation of state $w(z)$ in Figure \ref{F_w_DP+DD}. The points and their $1 \sigma$ error bars in this figure follow the same convention as in Figure \ref{F_H0_DPDD}. The yellow points indicate the value of $H_0$ with $1\sigma$ error obtained from gravitational lensing observations \citep{2020MNRAS.498.1420W,2020A&A...639A.101M,2023Sci...380.1322K,2025ApJ...979...13P}.}
    \label{F_H0_len+D+P}
\end{figure}

The descending trend of $H_0(z)$ with low significant level ($\sim2\sigma$) is found by the observations from strong gravitational lensing \citep{2020MNRAS.498.1420W,2020A&A...639A.101M,2023Sci...380.1322K,2025ApJ...979...13P}. In principle, the time delays between the images of a strongly lensed objects directly measure $H_0(z)$. We compile the $H_0(z)$ values derived from strong lensing observations, as shown in Figure \ref{F_H0_len+D+P}. It must be noted that the lens redshift is used \citep{2020MNRAS.498.1420W,2020A&A...639A.101M}. However, the constraint is from the time-delay distance, which depends on the combination of both the lens and source redshifts. It is important to note that the value of $H_0$ derived from strong gravitational lensing carries systematic uncertainties, as the constraints depends critically on the model of the lens galaxy.

In summary, all datasets - whether analyzed through joint constraints or individual probes - consistently demonstrate a concordant descending trend in $H_0(z)$. At low redshifts, the results are consistent with the SH0ES measurement, while they decline to match the Planck constraint at high redshifts. The evolution of the Hubble constant, as revealed by evolving dark energy, suggest that the Hubble tension - the discrepancy between early and late universe measurements - may originate from dynamical dark energy.

\section{Summary}\label{sec:summary}
To conclude, we consistently derived the evolution of $w(z)$ with redshift using a non-parametric method from DESI DR2 BAO measurements and SNe Ia. Although the absolute values show slight variations depending on the sample selection, all results remain consistent within $2\sigma$ confidence level. Our results show excellent agreement with the findings from DESI collaboration, further strengthening the robustness of the inferred evolution of $w(z)$. Notably, we derived the relation between the Hubble constant $H_0$ and the equation of state $w(z)$ of dark energy model-independently. From equation (\ref{eq_H0z}), the descending trend of $H_0(z)$ is found. This trend is robust and stable under different modeling choices. It agrees with that measured from the local distance ladder in $1\sigma$ confidence level at low redshift, and it gradually drops to the value measured from the CMB at high redshift. 

There are many different methods for measuring the value of the Hubble constant \citep{Yu2018,2022MNRAS.513.5686C,2022MNRAS.515L...1W,2024ApJ...977..120R,Gao2025,2025ApJ...985..203F,2025ApJ...981....9W}.
A marginal evidence on the evolution of
the Hubble constant was found before \citep{2020PhRvD.102b3520K, 2022MNRAS.517..576H,2023Univ....9...94H,Jia2023}. In the flat $\Lambda$CDM model, a descending trend in the Hubble constant has been found \citep{2020PhRvD.102j3525K,2020MNRAS.498.1420W,2021PhRvD.103j3509K,2021ApJ...912..150D,2022Galax..10...24D,2023arXiv230110572D,2024PDU....4401464O,2025JHEAp..4800405D,2025PhRvD.111h3547M,2025PDU....4801847M,Kalita2025}. However, the correlations between redshift bins are not removed in these analysis. After removing the correlations, the Hubble constant demonstrates redshift dependent evolutionary characteristics \citep{Jia2023,Jia2025}. Using model-independent Gaussian processes, a dip of $H_0$ around redshift $z \approx 0.5$ is found \citep{2022MNRAS.517..576H,2025MNRAS.542.1063H}. The evolution of $H_0$ may offer a compelling resolution to the Hubble tension, although the underlying physical mechanism remains unclear. If the actual universe consists of dynamical dark energy with varying $w(z)$, yet we assume it to be constant, this would lead to redshift-dependent variations in the $H_0$ values fitted under the $\Lambda$CDM model. 
Our result demonstrates that the descending trend is not coincidental, but rather stems from an underlying common physical mechanism. In future, the observations will yield deeper insights into the high-redshift universe, such as the China Space Station Telescope \citep{CSSTCollaboration2025}, Euclid \citep{EuclidCollaboration2025} and Rubin Observatory Legacy Survey of Space and Time \citep{Ivezic2019}. 

\section*{Acknowledgments}
We thank the anonymous referee for the helpful comments. This work was supported by the National Natural Science Foundation of China (grant Nos. 12494575 and 12273009).


\bibliography{ms}{}

@ARTICLE{Gao2025,
       author = {{Gao}, D.~H. and {Wu}, Q. and {Hu}, J.~P. and {Yi}, S.~X. and {Zhou}, X. and {Wang}, F.~Y. and {Dai}, Z.~G.},
        title = "{Measuring the Hubble constant using localized and nonlocalized fast radio bursts}",
      journal = {\aap},
         year = 2025,
        month = jun,
       volume = {698},
          eid = {A215},
        pages = {A215},
          doi = {10.1051/0004-6361/202453006},
archivePrefix = {arXiv},
       eprint = {2410.03994},
 primaryClass = {astro-ph.CO},
       adsurl = {https://ui.adsabs.harvard.edu/abs/2025A&A...698A.215G}
}

@ARTICLE{Freedman2021,
       author = {{Freedman}, Wendy L.},
        title = "{Measurements of the Hubble Constant: Tensions in Perspective}",
      journal = {\apj},
         year = 2021,
        month = sep,
       volume = {919},
       number = {1},
          eid = {16},
        pages = {16},
          doi = {10.3847/1538-4357/ac0e95},
archivePrefix = {arXiv},
       eprint = {2106.15656},
 primaryClass = {astro-ph.CO},
       adsurl = {https://ui.adsabs.harvard.edu/abs/2021ApJ...919...16F}
}

@ARTICLE{Li2025,
       author = {{Li}, Jun-Xian and {Wang}, Shuang},
        title = "{Reconstructing dark energy with model independent methods after DESI DR2 BAO}",
      journal = {arXiv e-prints},
         year = 2025,
        month = jun,
          eid = {arXiv:2506.22953},
        pages = {arXiv:2506.22953},
archivePrefix = {arXiv},
       eprint = {2506.22953},
 primaryClass = {astro-ph.CO},
       adsurl = {https://ui.adsabs.harvard.edu/abs/2025arXiv250622953L}
}

@ARTICLE{CSSTCollaboration2025,
       author = {{CSST Collaboration} and {Gong}, Yan and {Miao}, Haitao and {Zhan}, Hu and {Zu}, Ying},
        title = "{Introduction to the China Space Station Telescope (CSST)}",
      journal = {arXiv e-prints},
         year = 2025,
        month = jul,
          eid = {arXiv:2507.04618},
        pages = {arXiv:2507.04618},
archivePrefix = {arXiv},
       eprint = {2507.04618},
 primaryClass = {astro-ph.IM},
       adsurl = {https://ui.adsabs.harvard.edu/abs/2025arXiv250704618C}
}

@ARTICLE{EuclidCollaboration2025,
       author = {{Euclid Collaboration} and {Mellier}, Y. and {Abdurro'uf} and {Carrilho}, P. and {Carron Duque}, J. and {Carry}, B.},
        title = "{Euclid: I. Overview of the Euclid mission}",
      journal = {\aap},
         year = 2025,
        month = may,
       volume = {697},
          eid = {A1},
        pages = {A1},
          doi = {10.1051/0004-6361/202450810},
archivePrefix = {arXiv},
       eprint = {2405.13491},
 primaryClass = {astro-ph.CO},
       adsurl = {https://ui.adsabs.harvard.edu/abs/2025A&A...697A...1E}
}

@ARTICLE{Kalita2025,
       author = {{Kalita}, Surajit and {Uniyal}, Akhil and {Bulik}, Tomasz and {Mizuno}, Yosuke},
        title = "{Revealing Limitation in the Standard Cosmological Model: A Redshift-Dependent Hubble Constant from Fast Radio Bursts}",
      journal = {arXiv e-prints},
         year = 2025,
        month = jun,
          eid = {arXiv:2506.14947},
        pages = {arXiv:2506.14947},
          doi = {10.48550/arXiv.2506.14947},
archivePrefix = {arXiv},
       eprint = {2506.14947},
 primaryClass = {astro-ph.CO},
       adsurl = {https://ui.adsabs.harvard.edu/abs/2025arXiv250614947K}
}

@ARTICLE{Ivezic2019,
       author = {{Ivezi{\'c}}, {\v{Z}}eljko and {Kahn}, Steven M. and {Tyson}, J. Anthony and {Abel}, Bob and {Acosta}, Emily and {Allsman}, Robyn and {Alonso}, David and {AlSayyad}, Yusra and {Anderson}, Scott F. and {Andrew}, John and {Angel}, James Roger P. and {Angeli}, George Z. and {Ansari}, Reza and {Antilogus}, Pierre and {Araujo}, Constanza and {Armstrong}, Robert and {Arndt}, Kirk T. and {Astier}, Pierre and {Aubourg}, {\'E}ric and {Auza}, Nicole and {Axelrod}, Tim S. and {Bard}, Deborah J. and {Barr}, Jeff D. and {Barrau}, Aurelian and {Bartlett}, James G. and {Bauer}, Amanda E. and {Bauman}, Brian J. and {Baumont}, Sylvain and {Bechtol}, Ellen and {Bechtol}, Keith and {Becker}, Andrew C. and {Becla}, Jacek and {Beldica}, Cristina and {Bellavia}, Steve and {Bianco}, Federica B. and {Biswas}, Rahul and {Blanc}, Guillaume and {Blazek}, Jonathan and {Blandford}, Roger D. and {Bloom}, Josh S. and {Bogart}, Joanne and {Bond}, Tim W. and {Booth}, Michael T. and {Borgland}, Anders W. and {Borne}, Kirk and {Bosch}, James F. and {Boutigny}, Dominique and {Brackett}, Craig A. and {Bradshaw}, Andrew and {Brandt}, William Nielsen and {Brown}, Michael E. and {Bullock}, James S. and {Burchat}, Patricia and {Burke}, David L. and {Cagnoli}, Gianpietro and {Calabrese}, Daniel and {Callahan}, Shawn and {Callen}, Alice L. and {Carlin}, Jeffrey L. and {Carlson}, Erin L. and {Chandrasekharan}, Srinivasan and {Charles-Emerson}, Glenaver and {Chesley}, Steve and {Cheu}, Elliott C. and {Chiang}, Hsin-Fang and {Chiang}, James and {Chirino}, Carol and {Chow}, Derek and {Ciardi}, David R. and {Claver}, Charles F. and {Cohen-Tanugi}, Johann and {Cockrum}, Joseph J. and {Coles}, Rebecca and {Connolly}, Andrew J. and {Cook}, Kem H. and {Cooray}, Asantha and {Covey}, Kevin R. and {Cribbs}, Chris and {Cui}, Wei and {Cutri}, Roc and {Daly}, Philip N. and {Daniel}, Scott F. and {Daruich}, Felipe and {Daubard}, Guillaume and {Daues}, Greg and {Dawson}, William and {Delgado}, Francisco and {Dellapenna}, Alfred and {de Peyster}, Robert and {de Val-Borro}, Miguel and {Digel}, Seth W. and {Doherty}, Peter and {Dubois}, Richard and {Dubois-Felsmann}, Gregory P. and {Durech}, Josef and {Economou}, Frossie and {Eifler}, Tim and {Eracleous}, Michael and {Emmons}, Benjamin L. and {Fausti Neto}, Angelo and {Ferguson}, Henry and {Figueroa}, Enrique and {Fisher-Levine}, Merlin and {Focke}, Warren and {Foss}, Michael D. and {Frank}, James and {Freemon}, Michael D. and {Gangler}, Emmanuel and {Gawiser}, Eric and {Geary}, John C. and {Gee}, Perry and {Geha}, Marla and {Gessner}, Charles J.~B. and {Gibson}, Robert R. and {Gilmore}, D. Kirk and {Glanzman}, Thomas and {Glick}, William and {Goldina}, Tatiana and {Goldstein}, Daniel A. and {Goodenow}, Iain and {Graham}, Melissa L. and {Gressler}, William J. and {Gris}, Philippe and {Guy}, Leanne P. and {Guyonnet}, Augustin and {Haller}, Gunther and {Harris}, Ron and {Hascall}, Patrick A. and {Haupt}, Justine and {Hernandez}, Fabio and {Herrmann}, Sven and {Hileman}, Edward and {Hoblitt}, Joshua and {Hodgson}, John A. and {Hogan}, Craig and {Howard}, James D. and {Huang}, Dajun and {Huffer}, Michael E. and {Ingraham}, Patrick and {Innes}, Walter R. and {Jacoby}, Suzanne H. and {Jain}, Bhuvnesh and {Jammes}, Fabrice and {Jee}, M. James and {Jenness}, Tim and {Jernigan}, Garrett and {Jevremovi{\'c}}, Darko and {Johns}, Kenneth and {Johnson}, Anthony S. and {Johnson}, Margaret W.~G. and {Jones}, R. Lynne and {Juramy-Gilles}, Claire and {Juri{\'c}}, Mario and {Kalirai}, Jason S. and {Kallivayalil}, Nitya J. and {Kalmbach}, Bryce and {Kantor}, Jeffrey P. and {Karst}, Pierre and {Kasliwal}, Mansi M. and {Kelly}, Heather and {Kessler}, Richard and {Kinnison}, Veronica and {Kirkby}, David and {Knox}, Lloyd and {Kotov}, Ivan V. and {Krabbendam}, Victor L. and {Krughoff}, K. Simon and {Kub{\'a}nek}, Petr and {Kuczewski}, John and {Kulkarni}, Shri and {Ku}, John and {Kurita}, Nadine R. and {Lage}, Craig S. and {Lambert}, Ron and {Lange}, Travis and {Langton}, J. Brian and {Le Guillou}, Laurent and {Levine}, Deborah and {Liang}, Ming and {Lim}, Kian-Tat and {Lintott}, Chris J. and {Long}, Kevin E. and {Lopez}, Margaux and {Lotz}, Paul J. and {Lupton}, Robert H. and {Lust}, Nate B. and {MacArthur}, Lauren A. and {Mahabal}, Ashish and {Mandelbaum}, Rachel and {Markiewicz}, Thomas W. and {Marsh}, Darren S. and {Marshall}, Philip J. and {Marshall}, Stuart and {May}, Morgan and {McKercher}, Robert and {McQueen}, Michelle and {Meyers}, Joshua and {Migliore}, Myriam and {Miller}, Michelle and {Mills}, David J.},
        title = "{LSST: From Science Drivers to Reference Design and Anticipated Data Products}",
      journal = {\apj},
         year = 2019,
        month = mar,
       volume = {873},
       number = {2},
          eid = {111},
        pages = {111},
          doi = {10.3847/1538-4357/ab042c},
archivePrefix = {arXiv},
       eprint = {0805.2366},
 primaryClass = {astro-ph},
       adsurl = {https://ui.adsabs.harvard.edu/abs/2019ApJ...873..111I}
}

@ARTICLE{Peebles2003,
       author = {{Peebles}, P.~J. and {Ratra}, Bharat},
        title = "{The cosmological constant and dark energy}",
      journal = {Reviews of Modern Physics},
         year = 2003,
        month = apr,
       volume = {75},
       number = {2},
        pages = {559-606},
          doi = {10.1103/RevModPhys.75.559},
archivePrefix = {arXiv},
       eprint = {astro-ph/0207347},
 primaryClass = {astro-ph},
       adsurl = {https://ui.adsabs.harvard.edu/abs/2003RvMP...75..559P}
}

@ARTICLE{Perivolaropoulos2022,
       author = {{Perivolaropoulos}, L. and {Skara}, F.},
        title = "{Challenges for {\ensuremath{\Lambda}}CDM: An update}",
      journal = {\nar},
         year = 2022,
        month = dec,
       volume = {95},
          eid = {101659},
        pages = {101659},
          doi = {10.1016/j.newar.2022.101659},
archivePrefix = {arXiv},
       eprint = {2105.05208},
 primaryClass = {astro-ph.CO},
       adsurl = {https://ui.adsabs.harvard.edu/abs/2022NewAR..9501659P}
}

@ARTICLE{Benevento2020,
       author = {{Benevento}, Giampaolo and {Hu}, Wayne and {Raveri}, Marco},
        title = "{Can late dark energy transitions raise the Hubble constant?}",
      journal = {\prd},
         year = 2020,
        month = may,
       volume = {101},
       number = {10},
          eid = {103517},
        pages = {103517},
          doi = {10.1103/PhysRevD.101.103517},
archivePrefix = {arXiv},
       eprint = {2002.11707},
 primaryClass = {astro-ph.CO},
       adsurl = {https://ui.adsabs.harvard.edu/abs/2020PhRvD.101j3517B}
}

@ARTICLE{2022JHEAp..34...49A,
       author = {{Abdalla}, Elcio and {Abell{\'a}n}, Guillermo Franco and {Aboubrahim}, Amin and {Agnello}, Adriano and {Akarsu}, {\"O}zg{\"u}r and {Akrami}, Yashar and {Alestas}, George and {Aloni}, Daniel and {Amendola}, Luca and {Anchordoqui}, Luis A. and {Anderson}, Richard I. and {Arendse}, Nikki and {Asgari}, Marika and {Ballardini}, Mario and {Barger}, Vernon and {Basilakos}, Spyros and {Batista}, Ronaldo C. and {Battistelli}, Elia S. and {Battye}, Richard and {Benetti}, Micol and {Benisty}, David and {Berlin}, Asher and {de Bernardis}, Paolo and {Berti}, Emanuele and {Bidenko}, Bohdan and {Birrer}, Simon and {Blakeslee}, John P. and {Boddy}, Kimberly K. and {Bom}, Clecio R. and {Bonilla}, Alexander and {Borghi}, Nicola and {Bouchet}, Fran{\c{c}}ois R. and {Braglia}, Matteo and {Buchert}, Thomas and {Buckley-Geer}, Elizabeth and {Calabrese}, Erminia and {Caldwell}, Robert R. and {Camarena}, David and {Capozziello}, Salvatore and {Casertano}, Stefano and {Chen}, Geoff C. -F. and {Chluba}, Jens and {Chen}, Angela and {Chen}, Hsin-Yu and {Chudaykin}, Anton and {Cicoli}, Michele and {Copi}, Craig J. and {Courbin}, Fred and {Cyr-Racine}, Francis-Yan and {Czerny}, Bo{\.z}ena and {Dainotti}, Maria and {D'Amico}, Guido and {Davis}, Anne-Christine and {de Cruz P{\'e}rez}, Javier and {de Haro}, Jaume and {Delabrouille}, Jacques and {Denton}, Peter B. and {Dhawan}, Suhail and {Dienes}, Keith R. and {Di Valentino}, Eleonora and {Du}, Pu and {Eckert}, Dominique and {Escamilla-Rivera}, Celia and {Fert{\'e}}, Agn{\`e}s and {Finelli}, Fabio and {Fosalba}, Pablo and {Freedman}, Wendy L. and {Frusciante}, Noemi and {Gazta{\~n}aga}, Enrique and {Giar{\`e}}, William and {Giusarma}, Elena and {G{\'o}mez-Valent}, Adri{\`a} and {Handley}, Will and {Harrison}, Ian and {Hart}, Luke and {Hazra}, Dhiraj Kumar and {Heavens}, Alan and {Heinesen}, Asta and {Hildebrandt}, Hendrik and {Hill}, J. Colin and {Hogg}, Natalie B. and {Holz}, Daniel E. and {Hooper}, Deanna C. and {Hosseininejad}, Nikoo and {Huterer}, Dragan and {Ishak}, Mustapha and {Ivanov}, Mikhail M. and {Jaffe}, Andrew H. and {Jang}, In Sung and {Jedamzik}, Karsten and {Jimenez}, Raul and {Joseph}, Melissa and {Joudaki}, Shahab and {Kamionkowski}, Marc and {Karwal}, Tanvi and {Kazantzidis}, Lavrentios and {Keeley}, Ryan E. and {Klasen}, Michael and {Komatsu}, Eiichiro and {Koopmans}, L{\'e}on V.~E. and {Kumar}, Suresh and {Lamagna}, Luca and {Lazkoz}, Ruth and {Lee}, Chung-Chi and {Lesgourgues}, Julien and {Levi Said}, Jackson and {Lewis}, Tiffany R. and {L'Huillier}, Benjamin and {Lucca}, Matteo and {Maartens}, Roy and {Macri}, Lucas M. and {Marfatia}, Danny and {Marra}, Valerio and {Martins}, Carlos J.~A.~P. and {Masi}, Silvia and {Matarrese}, Sabino and {Mazumdar}, Arindam and {Melchiorri}, Alessandro and {Mena}, Olga and {Mersini-Houghton}, Laura and {Mertens}, James and {Milakovi{\'c}}, Dinko and {Minami}, Yuto and {Miranda}, Vivian and {Moreno-Pulido}, Cristian and {Moresco}, Michele and {Mota}, David F. and {Mottola}, Emil and {Mozzon}, Simone and {Muir}, Jessica and {Mukherjee}, Ankan and {Mukherjee}, Suvodip and {Naselsky}, Pavel and {Nath}, Pran and {Nesseris}, Savvas and {Niedermann}, Florian and {Notari}, Alessio and {Nunes}, Rafael C. and {{\'O} Colg{\'a}in}, Eoin and {Owens}, Kayla A. and {{\"O}z{\"u}lker}, Emre and {Pace}, Francesco and {Paliathanasis}, Andronikos and {Palmese}, Antonella and {Pan}, Supriya and {Paoletti}, Daniela and {Perez Bergliaffa}, Santiago E. and {Perivolaropoulos}, Leandros and {Pesce}, Dominic W. and {Pettorino}, Valeria and {Philcox}, Oliver H.~E. and {Pogosian}, Levon and {Poulin}, Vivian and {Poulot}, Gaspard and {Raveri}, Marco and {Reid}, Mark J. and {Renzi}, Fabrizio and {Riess}, Adam G. and {Sabla}, Vivian I. and {Salucci}, Paolo and {Salzano}, Vincenzo and {Saridakis}, Emmanuel N. and {Sathyaprakash}, Bangalore S. and {Schmaltz}, Martin and {Sch{\"o}neberg}, Nils and {Scolnic}, Dan and {Sen}, Anjan A. and {Sehgal}, Neelima and {Shafieloo}, Arman and {Sheikh-Jabbari}, M.~M. and {Silk}, Joseph and {Silvestri}, Alessandra and {Skara}, Foteini and {Sloth}, Martin S. and {Soares-Santos}, Marcelle and {Sol{\`a} Peracaula}, Joan and {Songsheng}, Yu-Yang and {Soriano}, Jorge F. and {Staicova}, Denitsa and {Starkman}, Glenn D. and {Szapudi}, Istv{\'a}n and {Teixeira}, Elsa M. and {Thomas}, Brooks and {Treu}, Tommaso and {Trott}, Emery and {van de Bruck}, Carsten and {Vazquez}, J. Alberto and {Verde}, Licia and {Visinelli}, Luca and {Wang}, Deng and {Wang}, Jian-Min and {Wang}, Shao-Jiang and {Watkins}, Richard and {Watson}, Scott and {Webb}, John K. and {Weiner}, Neal and {Weltman}, Amanda and {Witte}, Samuel J. and {Wojtak}, Rados{\l}aw and {Yadav}, Anil Kumar},
        title = "{Cosmology intertwined: A review of the particle physics, astrophysics, and cosmology associated with the cosmological tensions and anomalies}",
      journal = {Journal of High Energy Astrophysics},
         year = 2022,
        month = jun,
       volume = {34},
        pages = {49-211},
          doi = {10.1016/j.jheap.2022.04.002},
archivePrefix = {arXiv},
       eprint = {2203.06142},
 primaryClass = {astro-ph.CO},
       adsurl = {https://ui.adsabs.harvard.edu/abs/2022JHEAp..34...49A}
}

@ARTICLE{Scolnic2022,
       author = {{Scolnic}, Dan and {Brout}, Dillon and {Carr}, Anthony and {Riess}, Adam G. and {Davis}, Tamara M. and {Dwomoh}, Arianna and {Jones}, David O. and {Ali}, Noor and {Charvu}, Pranav and {Chen}, Rebecca and {Peterson}, Erik R. and {Popovic}, Brodie and {Rose}, Benjamin M. and {Wood}, Charlotte M. and {Brown}, Peter J. and {Chambers}, Ken and {Coulter}, David A. and {Dettman}, Kyle G. and {Dimitriadis}, Georgios and {Filippenko}, Alexei V. and {Foley}, Ryan J. and {Jha}, Saurabh W. and {Kilpatrick}, Charles D. and {Kirshner}, Robert P. and {Pan}, Yen-Chen and {Rest}, Armin and {Rojas-Bravo}, Cesar and {Siebert}, Matthew R. and {Stahl}, Benjamin E. and {Zheng}, WeiKang},
        title = "{The Pantheon+ Analysis: The Full Data Set and Light-curve Release}",
      journal = {\apj},
         year = 2022,
        month = oct,
       volume = {938},
       number = {2},
          eid = {113},
        pages = {113},
          doi = {10.3847/1538-4357/ac8b7a},
archivePrefix = {arXiv},
       eprint = {2112.03863},
 primaryClass = {astro-ph.CO},
       adsurl = {https://ui.adsabs.harvard.edu/abs/2022ApJ...938..113S}
}

@ARTICLE{DESCollaboration2024,
       author = {{DES Collaboration} and {Abbott}, T.~M.~C. and {Acevedo}, M. and {Aguena}, M. and {Alarcon}, A. and {Allam}, S. and {Alves}, O. and {Amon}, A. and {Andrade-Oliveira}, F. and {Annis}, J. and {Armstrong}, P. and {Asorey}, J. and {Avila}, S. and {Bacon}, D. and {Bassett}, B.~A. and {Bechtol}, K. and {Bernardinelli}, P.~H. and {Bernstein}, G.~M. and {Bertin}, E. and {Blazek}, J. and {Bocquet}, S. and {Brooks}, D. and {Brout}, D. and {Buckley-Geer}, E. and {Burke}, D.~L. and {Camacho}, H. and {Camilleri}, R. and {Campos}, A. and {Carnero Rosell}, A. and {Carollo}, D. and {Carr}, A. and {Carretero}, J. and {Castander}, F.~J. and {Cawthon}, R. and {Chang}, C. and {Chen}, R. and {Choi}, A. and {Conselice}, C. and {Costanzi}, M. and {da Costa}, L.~N. and {Crocce}, M. and {Davis}, T.~M. and {DePoy}, D.~L. and {Desai}, S. and {Diehl}, H.~T. and {Dixon}, M. and {Dodelson}, S. and {Doel}, P. and {Doux}, C. and {Drlica-Wagner}, A. and {Elvin-Poole}, J. and {Everett}, S. and {Ferrero}, I. and {Fert{\'e}}, A. and {Flaugher}, B. and {Foley}, R.~J. and {Fosalba}, P. and {Friedel}, D. and {Frieman}, J. and {Frohmaier}, C. and {Galbany}, L. and {Garc{\'\i}a-Bellido}, J. and {Gatti}, M. and {Gaztanaga}, E. and {Giannini}, G. and {Glazebrook}, K. and {Graur}, O. and {Gruen}, D. and {Gruendl}, R.~A. and {Gutierrez}, G. and {Hartley}, W.~G. and {Herner}, K. and {Hinton}, S.~R. and {Hollowood}, D.~L. and {Honscheid}, K. and {Huterer}, D. and {Jain}, B. and {James}, D.~J. and {Jeffrey}, N. and {Kasai}, E. and {Kelsey}, L. and {Kent}, S. and {Kessler}, R. and {Kim}, A.~G. and {Kirshner}, R.~P. and {Kovacs}, E. and {Kuehn}, K. and {Lahav}, O. and {Lee}, J. and {Lee}, S. and {Lewis}, G.~F. and {Li}, T.~S. and {Lidman}, C. and {Lin}, H. and {Malik}, U. and {Marshall}, J.~L. and {Martini}, P. and {Mena-Fern{\'a}ndez}, J. and {Menanteau}, F. and {Miquel}, R. and {Mohr}, J.~J. and {Mould}, J. and {Muir}, J. and {M{\"o}ller}, A. and {Neilsen}, E. and {Nichol}, R.~C. and {Nugent}, P. and {Ogando}, R.~L.~C. and {Palmese}, A. and {Pan}, Y. -C. and {Paterno}, M. and {Percival}, W.~J. and {Pereira}, M.~E.~S. and {Pieres}, A. and {Malag{\'o}n}, A.~A. Plazas and {Popovic}, B. and {Porredon}, A. and {Prat}, J. and {Qu}, H. and {Raveri}, M. and {Rodr{\'\i}guez-Monroy}, M. and {Romer}, A.~K. and {Roodman}, A. and {Rose}, B. and {Sako}, M. and {Sanchez}, E. and {Sanchez Cid}, D. and {Schubnell}, M. and {Scolnic}, D. and {Sevilla-Noarbe}, I. and {Shah}, P. and {Smith}, J. Allyn. and {Smith}, M. and {Soares-Santos}, M. and {Suchyta}, E. and {Sullivan}, M. and {Suntzeff}, N. and {Swanson}, M.~E.~C. and {S{\'a}nchez}, B.~O. and {Tarle}, G. and {Taylor}, G. and {Thomas}, D. and {To}, C. and {Toy}, M. and {Troxel}, M.~A. and {Tucker}, B.~E. and {Tucker}, D.~L. and {Uddin}, S.~A. and {Vincenzi}, M. and {Walker}, A.~R. and {Weaverdyck}, N. and {Wechsler}, R.~H. and {Weller}, J. and {Wester}, W. and {Wiseman}, P. and {Yamamoto}, M. and {Yuan}, F. and {Zhang}, B. and {Zhang}, Y.},
        title = "{The Dark Energy Survey: Cosmology Results with {\ensuremath{\sim}}1500 New High-redshift Type Ia Supernovae Using the Full 5 yr Data Set}",
      journal = {\apjl},
         year = 2024,
        month = sep,
       volume = {973},
       number = {1},
          eid = {L14},
        pages = {L14},
          doi = {10.3847/2041-8213/ad6f9f},
archivePrefix = {arXiv},
       eprint = {2401.02929},
 primaryClass = {astro-ph.CO},
       adsurl = {https://ui.adsabs.harvard.edu/abs/2024ApJ...973L..14D}
}

@ARTICLE{DiValentino2021,
       author = {{Di Valentino}, Eleonora and {Mena}, Olga and {Pan}, Supriya and {Visinelli}, Luca and {Yang}, Weiqiang and {Melchiorri}, Alessandro and {Mota}, David F. and {Riess}, Adam G. and {Silk}, Joseph},
        title = "{In the realm of the Hubble tension-a review of solutions}",
      journal = {Classical and Quantum Gravity},
         year = 2021,
        month = jul,
       volume = {38},
       number = {15},
          eid = {153001},
        pages = {153001},
          doi = {10.1088/1361-6382/ac086d},
archivePrefix = {arXiv},
       eprint = {2103.01183},
 primaryClass = {astro-ph.CO},
       adsurl = {https://ui.adsabs.harvard.edu/abs/2021CQGra..38o3001D}
}

@ARTICLE{Mortsell2022,
       author = {{M{\"o}rtsell}, Edvard and {Goobar}, Ariel and {Johansson}, Joel and {Dhawan}, Suhail},
        title = "{The Hubble Tension Revisited: Additional Local Distance Ladder Uncertainties}",
      journal = {\apj},
         year = 2022,
        month = aug,
       volume = {935},
       number = {1},
          eid = {58},
        pages = {58},
          doi = {10.3847/1538-4357/ac7c19},
archivePrefix = {arXiv},
       eprint = {2106.09400},
 primaryClass = {astro-ph.CO},
       adsurl = {https://ui.adsabs.harvard.edu/abs/2022ApJ...935...58M}
}

@ARTICLE{Chevallier2001,
       author = {{Chevallier}, Michel and {Polarski}, David},
        title = "{Accelerating Universes with Scaling Dark Matter}",
      journal = {International Journal of Modern Physics D},
         year = 2001,
        month = jan,
       volume = {10},
       number = {2},
        pages = {213-223},
          doi = {10.1142/S0218271801000822},
archivePrefix = {arXiv},
       eprint = {gr-qc/0009008},
 primaryClass = {gr-qc},
       adsurl = {https://ui.adsabs.harvard.edu/abs/2001IJMPD..10..213C}
}

@ARTICLE{Linder2003,
       author = {{Linder}, Eric V.},
        title = "{Exploring the Expansion History of the Universe}",
      journal = {\prl},
         year = 2003,
        month = mar,
       volume = {90},
       number = {9},
          eid = {091301},
        pages = {091301},
          doi = {10.1103/PhysRevLett.90.091301},
archivePrefix = {arXiv},
       eprint = {astro-ph/0208512},
 primaryClass = {astro-ph},
       adsurl = {https://ui.adsabs.harvard.edu/abs/2003PhRvL..90i1301L}
}

@ARTICLE{Adame2025,
       author = {{Adame}, A.~G. and {Aguilar}, J. and {Ahlen}, S. and {Alam}, S. and {Alexander}, D.~M. and {Alvarez}, M. and {Zhuang}, T.},
        title = "{DESI 2024 VI: cosmological constraints from the measurements of baryon acoustic oscillations}",
      journal = {\jcap},
         year = 2025,
        month = feb,
       volume = {2025},
       number = {2},
          eid = {021},
        pages = {021},
          doi = {10.1088/1475-7516/2025/02/021},
archivePrefix = {arXiv},
       eprint = {2404.03002},
 primaryClass = {astro-ph.CO},
       adsurl = {https://ui.adsabs.harvard.edu/abs/2025JCAP...02..021A}
}

@ARTICLE{Gu2025,
       author = {{Gu}, Gan and {Wang}, Xiaoma and {Wang}, Yuting and {Zhao}, Gong-Bo and {Pogosian}, Levon and {Koyama}, Kazuya and {Peacock}, John A. and {Cai}, Zheng and {Cervantes-Cota}, Jorge L. and {Zhao}, Ruiyang and {Ahlen}, Steven and {Bianchi}, Davide and {Brooks}, David and {Claybaugh}, Todd and {Cole}, Shaun and {de la Macorra}, Axel and {de Mattia}, Arnaud and {Doel}, Peter and {Ferraro}, Simone and {Forero-Romero}, Jaime E. and {Gazta{\~n}aga}, Enrique and {Gontcho}, Satya Gontcho A and {Gutierrez}, Gaston and {Hahn}, ChangHoon and {Howlett}, Cullan and {Ishak}, Mustapha and {Kehoe}, Robert and {Kirkby}, David and {Kneib}, Jean-Paul and {Lahav}, Ofer and {Landriau}, Martin and {Le Guillou}, Laurent and {Leauthaud}, Alexie and {Levi}, Michael and {Manera}, Marc and {Meisner}, Aaron and {Miquel}, Ramon and {Moustakas}, John and {Mu{\~n}oz-Guti{\'e}rrez}, Andrea and {Nadathur}, Seshadri and {Newman}, Jeffrey A. and {Palanque-Delabrouille}, Nathalie and {Percival}, Will and {Prada}, Francisco and {P{\'e}rez-R{\`a}fols}, Ignasi and {Rossi}, Graziano and {Samushia}, Lado and {Sanchez}, Eusebio and {Schlegel}, David and {Seo}, Hee-Jong and {Shafieloo}, Arman and {Sprayberry}, David and {Tarl{\'e}}, Gregory and {Walther}, Michael and {Weaver}, Benjamin Alan and {Zarrouk}, Pauline and {Zhao}, Cheng and {Zhou}, Rongpu and {Zou}, Hu},
        title = "{Dynamical Dark Energy in light of the DESI DR2 Baryonic Acoustic Oscillations Measurements}",
      journal = {arXiv e-prints},
         year = 2025,
        month = apr,
          eid = {arXiv:2504.06118},
        pages = {arXiv:2504.06118},
          doi = {10.48550/arXiv.2504.06118},
archivePrefix = {arXiv},
       eprint = {2504.06118},
 primaryClass = {astro-ph.CO},
       adsurl = {https://ui.adsabs.harvard.edu/abs/2025arXiv250406118G}
}

@ARTICLE{DESI_DR2,
       author = {{DESI Collaboration} and {Abdul-Karim}, M. and {Aguilar}, J. and {Ahlen}, S. and {Alam}, S. and {Allen}, L. and {Allende Prieto}, C. and {Alves}, O. and {Anand}, A. and {Zou}, H.},
        title = "{DESI DR2 Results II: Measurements of Baryon Acoustic Oscillations and Cosmological Constraints}",
      journal = {arXiv e-prints},
         year = 2025,
        month = mar,
          eid = {arXiv:2503.14738},
        pages = {arXiv:2503.14738},
          doi = {10.48550/arXiv.2503.14738},
archivePrefix = {arXiv},
       eprint = {2503.14738},
 primaryClass = {astro-ph.CO},
       adsurl = {https://ui.adsabs.harvard.edu/abs/2025arXiv250314738D}
}

@ARTICLE{Planck_2020,
       author = {{Planck Collaboration} and {Aghanim}, N. and {Akrami}, Y. and {Ashdown}, M. and {Aumont}, J. and {Baccigalupi}, C. and {Ballardini}, M. and {Banday}, A.~J. and {Barreiro}, R.~B. and {Bartolo}, N. and {Basak}, S. and {Battye}, R. and {Benabed}, K. and {Bernard}, J. -P. and {Bersanelli}, M. and {Bielewicz}, P. and {Bock}, J.~J. and {Bond}, J.~R. and {Borrill}, J. and {Bouchet}, F.~R. and {Boulanger}, F. and {Bucher}, M. and {Burigana}, C. and {Butler}, R.~C. and {Calabrese}, E. and {Cardoso}, J. -F. and {Carron}, J. and {Challinor}, A. and {Chiang}, H.~C. and {Chluba}, J. and {Colombo}, L.~P.~L. and {Combet}, C. and {Contreras}, D. and {Crill}, B.~P. and {Cuttaia}, F. and {de Bernardis}, P. and {de Zotti}, G. and {Delabrouille}, J. and {Delouis}, J. -M. and {Di Valentino}, E. and {Diego}, J.~M. and {Dor{\'e}}, O. and {Douspis}, M. and {Ducout}, A. and {Dupac}, X. and {Dusini}, S. and {Efstathiou}, G. and {Elsner}, F. and {En{\ss}lin}, T.~A. and {Eriksen}, H.~K. and {Fantaye}, Y. and {Farhang}, M. and {Fergusson}, J. and {Fernandez-Cobos}, R. and {Finelli}, F. and {Forastieri}, F. and {Frailis}, M. and {Fraisse}, A.~A. and {Franceschi}, E. and {Frolov}, A. and {Galeotta}, S. and {Galli}, S. and {Ganga}, K. and {G{\'e}nova-Santos}, R.~T. and {Gerbino}, M. and {Ghosh}, T. and {Gonz{\'a}lez-Nuevo}, J. and {G{\'o}rski}, K.~M. and {Gratton}, S. and {Gruppuso}, A. and {Gudmundsson}, J.~E. and {Hamann}, J. and {Handley}, W. and {Hansen}, F.~K. and {Herranz}, D. and {Hildebrandt}, S.~R. and {Hivon}, E. and {Huang}, Z. and {Jaffe}, A.~H. and {Jones}, W.~C. and {Karakci}, A. and {Keih{\"a}nen}, E. and {Keskitalo}, R. and {Kiiveri}, K. and {Kim}, J. and {Kisner}, T.~S. and {Knox}, L. and {Krachmalnicoff}, N. and {Kunz}, M. and {Kurki-Suonio}, H. and {Lagache}, G. and {Lamarre}, J. -M. and {Lasenby}, A. and {Lattanzi}, M. and {Lawrence}, C.~R. and {Le Jeune}, M. and {Lemos}, P. and {Lesgourgues}, J. and {Levrier}, F. and {Lewis}, A. and {Liguori}, M. and {Lilje}, P.~B. and {Lilley}, M. and {Lindholm}, V. and {L{\'o}pez-Caniego}, M. and {Lubin}, P.~M. and {Ma}, Y. -Z. and {Mac{\'\i}as-P{\'e}rez}, J.~F. and {Maggio}, G. and {Maino}, D. and {Mandolesi}, N. and {Mangilli}, A. and {Marcos-Caballero}, A. and {Maris}, M. and {Martin}, P.~G. and {Martinelli}, M. and {Mart{\'\i}nez-Gonz{\'a}lez}, E. and {Matarrese}, S. and {Mauri}, N. and {McEwen}, J.~D. and {Meinhold}, P.~R. and {Melchiorri}, A. and {Mennella}, A. and {Migliaccio}, M. and {Millea}, M. and {Mitra}, S. and {Miville-Desch{\^e}nes}, M. -A. and {Molinari}, D. and {Montier}, L. and {Morgante}, G. and {Moss}, A. and {Natoli}, P. and {N{\o}rgaard-Nielsen}, H.~U. and {Pagano}, L. and {Paoletti}, D. and {Partridge}, B. and {Patanchon}, G. and {Peiris}, H.~V. and {Perrotta}, F. and {Pettorino}, V. and {Piacentini}, F. and {Polastri}, L. and {Polenta}, G. and {Puget}, J. -L. and {Rachen}, J.~P. and {Reinecke}, M. and {Remazeilles}, M. and {Renzi}, A. and {Rocha}, G. and {Rosset}, C. and {Roudier}, G. and {Rubi{\~n}o-Mart{\'\i}n}, J.~A. and {Ruiz-Granados}, B. and {Salvati}, L. and {Sandri}, M. and {Savelainen}, M. and {Scott}, D. and {Shellard}, E.~P.~S. and {Sirignano}, C. and {Sirri}, G. and {Spencer}, L.~D. and {Sunyaev}, R. and {Suur-Uski}, A. -S. and {Tauber}, J.~A. and {Tavagnacco}, D. and {Tenti}, M. and {Toffolatti}, L. and {Tomasi}, M. and {Trombetti}, T. and {Valenziano}, L. and {Valiviita}, J. and {Van Tent}, B. and {Vibert}, L. and {Vielva}, P. and {Villa}, F. and {Vittorio}, N. and {Wandelt}, B.~D. and {Wehus}, I.~K. and {White}, M. and {White}, S.~D.~M. and {Zacchei}, A. and {Zonca}, A.},
        title = "{Planck 2018 results. VI. Cosmological parameters}",
      journal = {\aap},
         year = 2020,
        month = sep,
       volume = {641},
          eid = {A6},
        pages = {A6},
          doi = {10.1051/0004-6361/201833910},
archivePrefix = {arXiv},
       eprint = {1807.06209},
 primaryClass = {astro-ph.CO},
       adsurl = {https://ui.adsabs.harvard.edu/abs/2020A&A...641A...6P}
}

@ARTICLE{Riess2022,
       author = {{Riess}, Adam G. and {Yuan}, Wenlong and {Macri}, Lucas M. and {Scolnic}, Dan and {Brout}, Dillon and {Casertano}, Stefano and {Jones}, David O. and {Murakami}, Yukei and {Anand}, Gagandeep S. and {Breuval}, Louise and {Brink}, Thomas G. and {Filippenko}, Alexei V. and {Hoffmann}, Samantha and {Jha}, Saurabh W. and {D'arcy Kenworthy}, W. and {Mackenty}, John and {Stahl}, Benjamin E. and {Zheng}, WeiKang},
        title = "{A Comprehensive Measurement of the Local Value of the Hubble Constant with 1 km s$^{-1}$ Mpc$^{-1}$ Uncertainty from the Hubble Space Telescope and the SH0ES Team}",
      journal = {\apjl},
         year = 2022,
        month = jul,
       volume = {934},
       number = {1},
          eid = {L7},
        pages = {L7},
          doi = {10.3847/2041-8213/ac5c5b},
archivePrefix = {arXiv},
       eprint = {2112.04510},
 primaryClass = {astro-ph.CO},
       adsurl = {https://ui.adsabs.harvard.edu/abs/2022ApJ...934L...7R}
}

@ARTICLE{Jia2022,
       author = {{Jia}, X.~D. and {Hu}, J.~P. and {Yang}, J. and {Zhang}, B.~B. and {Wang}, F.~Y.},
        title = "{E $_{iso}$-E$_{p}$ correlation of gamma-ray bursts: calibration and cosmological applications}",
      journal = {\mnras},
         year = 2022,
        month = oct,
       volume = {516},
       number = {2},
        pages = {2575-2585},
          doi = {10.1093/mnras/stac2356},
archivePrefix = {arXiv},
       eprint = {2208.09272},
 primaryClass = {astro-ph.HE},
       adsurl = {https://ui.adsabs.harvard.edu/abs/2022MNRAS.516.2575J}
}

@ARTICLE{Riess2007,
       author = {{Riess}, Adam G. and {Strolger}, Louis-Gregory and {Casertano}, Stefano and {Ferguson}, Henry C. and {Mobasher}, Bahram and {Gold}, Ben and {Challis}, Peter J. and {Filippenko}, Alexei V. and {Jha}, Saurabh and {Li}, Weidong and {Tonry}, John and {Foley}, Ryan and {Kirshner}, Robert P. and {Dickinson}, Mark and {MacDonald}, Emily and {Eisenstein}, Daniel and {Livio}, Mario and {Younger}, Josh and {Xu}, Chun and {Dahl{\'e}n}, Tomas and {Stern}, Daniel},
        title = "{New Hubble Space Telescope Discoveries of Type Ia Supernovae at z >= 1: Narrowing Constraints on the Early Behavior of Dark Energy}",
      journal = {\apj},
         year = 2007,
        month = apr,
       volume = {659},
       number = {1},
        pages = {98-121},
          doi = {10.1086/510378},
archivePrefix = {arXiv},
       eprint = {astro-ph/0611572},
 primaryClass = {astro-ph},
       adsurl = {https://ui.adsabs.harvard.edu/abs/2007ApJ...659...98R}
}

@ARTICLE{Huterer2005,
       author = {{Huterer}, Dragan and {Cooray}, Asantha},
        title = "{Uncorrelated estimates of dark energy evolution}",
      journal = {\prd},
         year = 2005,
        month = jan,
       volume = {71},
       number = {2},
          eid = {023506},
        pages = {023506},
          doi = {10.1103/PhysRevD.71.023506},
archivePrefix = {arXiv},
       eprint = {astro-ph/0404062},
 primaryClass = {astro-ph},
       adsurl = {https://ui.adsabs.harvard.edu/abs/2005PhRvD..71b3506H}
}

@ARTICLE{Jia2025,
       author = {{Jia}, X.~D. and {Hu}, J.~P. and {Yi}, S.~X. and {Wang}, F.~Y.},
        title = "{Uncorrelated Estimations of H$_{0}$ Redshift Evolution from DESI Baryon Acoustic Oscillation Observations}",
      journal = {\apjl},
         year = 2025,
        month = feb,
       volume = {979},
       number = {2},
          eid = {L34},
        pages = {L34},
          doi = {10.3847/2041-8213/ada94d},
archivePrefix = {arXiv},
       eprint = {2406.02019},
 primaryClass = {astro-ph.CO},
       adsurl = {https://ui.adsabs.harvard.edu/abs/2025ApJ...979L..34J}
}

@ARTICLE{2019NatAs...3..891V,
       author = {{Verde}, Licia and {Treu}, Tommaso and {Riess}, Adam G.},
        title = "{Tensions between the early and late Universe}",
      journal = {Nature Astronomy},
         year = 2019,
        month = sep,
       volume = {3},
        pages = {891-895},
          doi = {10.1038/s41550-019-0902-0},
archivePrefix = {arXiv},
       eprint = {1907.10625},
 primaryClass = {astro-ph.CO},
       adsurl = {https://ui.adsabs.harvard.edu/abs/2019NatAs...3..891V}
}

@ARTICLE{Riess2020,
       author = {{Riess}, Adam G.},
        title = "{The expansion of the Universe is faster than expected}",
      journal = {Nature Reviews Physics},
         year = 2020,
        month = jan,
       volume = {2},
       number = {1},
        pages = {10-12},
          doi = {10.1038/s42254-019-0137-0},
archivePrefix = {arXiv},
       eprint = {2001.03624},
 primaryClass = {astro-ph.CO},
       adsurl = {https://ui.adsabs.harvard.edu/abs/2020NatRP...2...10R}
}

@ARTICLE{2019PhRvL.122v1301P,
       author = {{Poulin}, Vivian and {Smith}, Tristan L. and {Karwal}, Tanvi and {Kamionkowski}, Marc},
        title = "{Early Dark Energy can Resolve the Hubble Tension}",
      journal = {\prl},
         year = 2019,
        month = jun,
       volume = {122},
       number = {22},
          eid = {221301},
        pages = {221301},
          doi = {10.1103/PhysRevLett.122.221301},
archivePrefix = {arXiv},
       eprint = {1811.04083},
 primaryClass = {astro-ph.CO},
       adsurl = {https://ui.adsabs.harvard.edu/abs/2019PhRvL.122v1301P}
}

@ARTICLE{2020PhRvL.125r1302J,
       author = {{Jedamzik}, Karsten and {Pogosian}, Levon},
        title = "{Relieving the Hubble Tension with Primordial Magnetic Fields}",
      journal = {\prl},
         year = 2020,
        month = oct,
       volume = {125},
       number = {18},
          eid = {181302},
        pages = {181302},
          doi = {10.1103/PhysRevLett.125.181302},
archivePrefix = {arXiv},
       eprint = {2004.09487},
 primaryClass = {astro-ph.CO},
       adsurl = {https://ui.adsabs.harvard.edu/abs/2020PhRvL.125r1302J}
}

@ARTICLE{2017NatAs...1..627Z,
       author = {{Zhao}, Gong-Bo and {Raveri}, Marco and {Pogosian}, Levon and {Wang}, Yuting and {Crittenden}, Robert G. and {Handley}, Will J. and {Percival}, Will J. and {Beutler}, Florian and {Brinkmann}, Jonathan and {Chuang}, Chia-Hsun and {Cuesta}, Antonio J. and {Eisenstein}, Daniel J. and {Kitaura}, Francisco-Shu and {Koyama}, Kazuya and {L'Huillier}, Benjamin and {Nichol}, Robert C. and {Pieri}, Matthew M. and {Rodriguez-Torres}, Sergio and {Ross}, Ashley J. and {Rossi}, Graziano and {S{\'a}nchez}, Ariel G. and {Shafieloo}, Arman and {Tinker}, Jeremy L. and {Tojeiro}, Rita and {Vazquez}, Jose A. and {Zhang}, Hanyu},
        title = "{Dynamical dark energy in light of the latest observations}",
      journal = {Nature Astronomy},
         year = 2017,
        month = aug,
       volume = {1},
        pages = {627-632},
          doi = {10.1038/s41550-017-0216-z},
archivePrefix = {arXiv},
       eprint = {1701.08165},
 primaryClass = {astro-ph.CO},
       adsurl = {https://ui.adsabs.harvard.edu/abs/2017NatAs...1..627Z}
}

@ARTICLE{2025MNRAS.536.3232M,
       author = {{Mazurenko}, Sergij and {Banik}, Indranil and {Kroupa}, Pavel},
        title = "{The redshift dependence of the inferred H$_{0}$ in a local void solution to the Hubble tension}",
      journal = {\mnras},
         year = 2025,
        month = feb,
       volume = {536},
       number = {4},
        pages = {3232-3241},
          doi = {10.1093/mnras/stae2758},
archivePrefix = {arXiv},
       eprint = {2412.12245},
 primaryClass = {astro-ph.CO},
       adsurl = {https://ui.adsabs.harvard.edu/abs/2025MNRAS.536.3232M}
}

@ARTICLE{Pantheon+,
       author = {{Brout}, Dillon and {Scolnic}, Dan and {Popovic}, Brodie and {Riess}, Adam G. and {Carr}, Anthony and {Zuntz}, Joe and {Kessler}, Rick and {Davis}, Tamara M. and {Hinton}, Samuel and {Jones}, David and {Kenworthy}, W. D'Arcy and {Peterson}, Erik R. and {Said}, Khaled and {Taylor}, Georgie and {Ali}, Noor and {Armstrong}, Patrick and {Charvu}, Pranav and {Dwomoh}, Arianna and {Meldorf}, Cole and {Palmese}, Antonella and {Qu}, Helen and {Rose}, Benjamin M. and {Sanchez}, Bruno and {Stubbs}, Christopher W. and {Vincenzi}, Maria and {Wood}, Charlotte M. and {Brown}, Peter J. and {Chen}, Rebecca and {Chambers}, Ken and {Coulter}, David A. and {Dai}, Mi and {Dimitriadis}, Georgios and {Filippenko}, Alexei V. and {Foley}, Ryan J. and {Jha}, Saurabh W. and {Kelsey}, Lisa and {Kirshner}, Robert P. and {M{\"o}ller}, Anais and {Muir}, Jessie and {Nadathur}, Seshadri and {Pan}, Yen-Chen and {Rest}, Armin and {Rojas-Bravo}, Cesar and {Sako}, Masao and {Siebert}, Matthew R. and {Smith}, Mat and {Stahl}, Benjamin E. and {Wiseman}, Phil},
        title = "{The Pantheon+ Analysis: Cosmological Constraints}",
      journal = {\apj},
         year = 2022,
        month = oct,
       volume = {938},
       number = {2},
          eid = {110},
        pages = {110},
          doi = {10.3847/1538-4357/ac8e04},
archivePrefix = {arXiv},
       eprint = {2202.04077},
 primaryClass = {astro-ph.CO},
       adsurl = {https://ui.adsabs.harvard.edu/abs/2022ApJ...938..110B}
}
\bibliographystyle{aasjournal}

\begin{table*}
	\centering
	\caption{Fitting results of the equation of state of dark energy $w(z)$ with different datasets. The best-fit values of $w(z)$ with $1 \sigma$ confidence level are given. \label{T}}
	\begin{tabular}{ccccc}
		\\
		\hline
		$w(z)$	  & DESI + Pantheon plus      & DESI + DESY5              & DESI                    & Pantheon plus       \\
		\hline
		$w_1$       & $-1.06^{+0.04}_{-0.04}$   & $-0.87^{+0.04}_{-0.04}$   & $-1.02^{+0.08}_{-0.08}$        & $-0.96^{+0.05}_{-0.05}$\\
		$w_2$       & $-1.25^{+0.06}_{-0.06}$   & $-1.13^{+0.06}_{-0.06}$   & $-0.71^{+0.13}_{-0.14}$        & $-0.92^{+0.09}_{-0.10}$\\
		$w_3$       & $-1.25^{+0.11}_{-0.12}$   & $-1.27^{+0.11}_{-0.12}$   & $-1.14^{+0.20}_{-0.20}$        & $-0.81^{+0.26}_{-0.30}$\\
		$w_4$       & $-1.38^{+0.22}_{-0.23}$   & $-1.46^{+0.23}_{-0.24}$   & $-1.90^{+0.57}_{-0.87}$        & $-2.59^{+0.84}_{-0.62}$\\
		$w_5$       & $-2.84^{+1.28}_{-1.89}$   & $-3.02^{+1.06}_{-1.36}$   & -                              & -                     \\
		\hline
	\end{tabular}
\end{table*}

\begin{table*}\large
	\caption{Results of the Hubble constant $H_0$ in units of km s$^{-1}$ Mpc$^{-1}$ with different datasets. The values of $H_0$ are calculated from the results of $w(z)$ by equation (\ref{eq_H0z}). The values of $H_{0, i}$ represent the maximum a posteriori estimate in each redshift bin, along with the $1 \sigma$ confidence level.  \label{T_H}}
	\centering
	\begin{tabular}{ccccc}
		\\
		\hline
		$H_0$	  & DESI + Pantheon plus      & DESI + DESY5              & DESI                    & Pantheon plus       \\
		\hline
		$H_{0, 1}$  & $72.20^{+0.19}_{-0.19}$   & $72.98^{+0.17}_{-0.17}$   & $69.98^{+0.99}_{-0.99}$        & $73.48^{+0.15}_{-0.15}$\\
		$H_{0, 2}$  & $71.62^{+0.36}_{-0.36}$   & $70.69^{+0.42}_{-0.42}$   & $70.63^{+1.16}_{-1.16}$        & $73.97^{+0.54}_{-0.54}$\\
		$H_{0, 3}$  & $69.78^{+0.51}_{-0.51}$   & $68.75^{+0.53}_{-0.53}$   & $70.51^{+1.27}_{-1.27}$        & $74.64^{+1.09}_{-1.09}$\\
		$H_{0, 4}$  & $68.13^{+0.67}_{-0.67}$   & $67.54^{+0.67}_{-0.67}$   & $68.12^{+0.71}_{-0.71}$        & $69.77^{+1.41}_{-1.41}$\\
		$H_{0, 5}$  & $67.23^{+0.84}_{-0.84}$   & $67.29^{+0.84}_{-0.84}$   & $66.59^{+0.78}_{-0.78}$        & -                        \\
		\hline
	\end{tabular}
\end{table*}

\end{document}